\documentclass[11pt,a4paper]{article}
\usepackage{jcappub}
\usepackage{bm}
\usepackage{subfigure}
\usepackage{setspace}

\newcommand{\alert}{\textcolor{black}}

\title{Cosmology with the largest galaxy cluster surveys: Going beyond Fisher matrix forecasts}

\author[]{Satej Khedekar$^{1}$ and}
\author[]{Subhabrata Majumdar$^{2}$}

\affiliation[]{$^1$MPI f\"ur Astrophysik, Karl-Schwarzschild str. 1, Garching, 85741, Germany\\
$^2$Tata Institute of Fundamental Research \\Homi Bhabha Road, Colaba, Mumbai - 400005, India.}
\emailAdd{satej@mpa-garching.mpg.de}
\emailAdd{subha@tifr.res.in}

\abstract{
We make the first detailed MCMC likelihood study of cosmological constraints that are expected from some of the largest, ongoing and proposed, cluster surveys in different wave-bands  and compare the estimates to the prevalent Fisher matrix forecasts. Mock catalogs of cluster counts
expected from the surveys -- eROSITA, WFXT, RCS2, DES and Planck, along with a mock dataset
of follow-up mass calibrations are analyzed for this purpose.
A fair agreement between MCMC and Fisher results is found
only in the case of minimal models. However, for many cases, the marginalized constraints obtained from 
Fisher and MCMC methods can differ by factors of 30-100\%.
The discrepancy can be alarmingly large for a time dependent dark energy equation of state, $w(a)$; 
the Fisher methods are seen to under-estimate the constraints by as much as a factor of 4--5. 
Typically, Fisher estimates become more and more inappropriate as we move away from $\Lambda$CDM, to a constant-$w$ dark energy to varying-$w$ dark energy cosmologies.
Fisher analysis, also, predicts incorrect parameter degeneracies. There are noticeable offsets in the likelihood contours obtained from 
Fisher methods that is caused due to an asymmetry in the posterior likelihood distribution as seen through a MCMC 
analysis. From the point of mass-calibration uncertainties, a
high value of unknown scatter about the mean mass-observable relation, and its redshift dependence, is seen to have large 
degeneracies with the cosmological parameters $\sigma_8$ and $w(a)$ and can degrade the cosmological constraints considerably. 
We find that the addition of mass-calibrated cluster datasets can improve dark energy and $\sigma_8$ constraints by factors of 2--3 from what can be obtained from CMB+SNe+BAO only .
Finally, we show that a joint analysis of datasets of two (or more) different cluster surveys would significantly tighten cosmological constraints from using clusters only. Since, details of future cluster surveys are still being planned, we emphasize that optimal survey design must be done using MCMC analysis rather than Fisher forecasting.
}


\arxivnumber{1210.5586}
  
\begin{document}

\maketitle
\flushbottom

\section{Introduction}
\label{sec:Introduction}
Galaxy cluster counts have been recognized as potentially important probes of precision cosmology
\cite{Holder01, Hu, Levine, WangSteinhardt, Weller02, Molnar}, and have yielded in recent years,
 the first set of independent constraints on $\Omega_m$, the matter density of the Universe and
 $\sigma_8$, the RMS density fluctuations on a scale of $8~h^{-1}$ Mpc \cite{ROSAT_cosmo,
 Rozo10, Chandra_cosmo, Gladders}. There are many large scale structure (LSS) surveys
 that are either operational or being built, which are expected to yield a much larger
 dataset of cluster observations. This would provide strong constraints not only on $\Omega_m$
 and $\sigma_8$, but also on the dark energy equation of state ($w_0$) and its evolution ($w_a / w_z$).
 There have also been recent works which demonstrate that cluster data can also be used to
 place constraints on the extensions of the standard cosmological model, for e.g. probing
 non-Gaussianities in primordial density perturbations \cite{Sartoris_ng, Cunha_ng, Fedeli_ng,
 Roncarelli_ng}, modifications to general relativity \cite{Rapetti_GR1, Rapetti_GR2} and to
 probe neutrino properties \cite{Mantz_neutrino}. 
The cosmological constraints obtained from galaxy clusters are complementary to those obtained
 from other cosmological probes such as cosmic microwave background (CMB),
 supernovae type Ia observations and baryon acoustic oscillations. This is because
 clusters have the dual advantage of not only probing the expansion history but also the
 growth of perturbations in the universe. Using datasets from various different probes helps in breaking
 degeneracies between various cosmological parameters, and gives tighter constraints
 \cite{Bahcall,Huterer,Levine,Battye,WangKhoury,Melchiorri,MantzAllen}.

In recent years there have been a large number of cluster surveys running, funded or proposed with the aim of detecting between thousand to hundreds of thousands of clusters up to high redshifts ($z\gtrsim1$) 
\footnote{www.darkenergysurvey.org \hspace*{0.5cm}
www.rcs2.org \hspace*{0.5cm}
www.mpe.mpg.de/eROSITA \hspace*{0.5cm}
wfxt.pha.jhu.edu \hspace*{0.5cm}
www.rssd.esa.int/index.php?project=planck \hspace*{0.5cm}
pole.uchicago.edu \hspace*{0.5cm}
www.princeton.edu/act}. 
 With the newer surveys being proposed, it is important to asses the cosmological relevance of such observations.
 Fisher based estimates of the constraints have been widely used in making forecasts because of both
 simplicity and lower computational costs \cite{Pillepich2012}. However, Fisher estimates are reliable only when the
 underlying probability distribution function (pdf) is Gaussian shaped, while MCMC forecasts are
 always correct irrespective of the pdf of the likelihood. MCMC techniques can be used on
 real data whereas Fisher analysis is limited to making forecasts. Detailed comparisons of 
forecasts from Fisher and MCMC have been made in the context of CMB in \cite{Perotto} and SNe, BAO and weak lensing in \cite{Wolz2012}. However, 
no such comparison has been made for cluster surveys and in this work, we make a first detailed comparison for cluster surveys. Unlike the CMB where there are a few 
nuisance parameters like $\tau$ (optical depth) and $Y_{\rm He}$ (Helium abundance), analysis 
of cluster data requires a larger number of such parameters to arrive at unbiased 
cosmological parameter values. This is due to the significant non-linear astrophysical processes 
occurring in the intra-cluster medium (ICM) (much of which are still not well understood),
 as compared to the CMB physics which is both linear and well understood.

In this paper, we compare results on parameter constraints from both Fisher and MCMC methods for
 various cluster surveys in x-rays, SZE and optical. We make detailed comparison on a case by case
 basis for various cosmological models like $\Lambda$CDM\footnote{Cold dark matter with the 
cosmological constant as the dark energy component.}, $w$CDM\footnote{Cold dark matter with a 
single parameter dark energy equation of state, $w_0$.} and also $w_a$CDM($w_z$CDM)\footnote{Cold 
dark matter with an evolving dark energy equation of state, described by two parameters -- $w_0$ 
and $w_a$($w_z$), see section \ref{sec:fiducial_model} for more details.}. The rest of the paper is 
organized as follows. In section \ref{sec:dndz} we begin by describing how clusters are used to
 probe cosmology. In section \ref{sec:MCMC} and \ref{sec:Fisher} we summarize the methodology of 
MCMC and Fisher analysis in the context of cluster data and also mention the advantages and 
disadvantages of each method. In section \ref{sec:fiducial_model} we describe our fiducial 
cosmological models, and the priors used in this work; we also outline here our procedure for 
constructing the mock mass follow up catalog. Next, in section \ref{sec:surveys} we move on to
 briefly describe the six surveys across multiple wavelength bands that have been considered in the paper.
 Our results are described in detail in Section \ref{sec:results} and are also highlighted in 
Tables \ref{tab:eROSITA_results} - \ref{tab:CMB_priors}, which also offer a detailed comparison
 between MCMC and Fisher forecasts. Here, we also provide some examples of how the 
synergies of two independent cluster surveys may be utilized to break the parameter degeneracies 
to obtain significantly better cosmological constraints. We discuss some technical points relevant 
to our work in section \ref{sec:discussions} before concluding in section \ref{sec:conclusion}.

\section{Redshift distribution of clusters as a probe of cosmology}
\label{sec:dndz}
The redshift distribution of clusters is given by,
\begin{equation}
\frac{dN}{dz}(z)=\Delta\Omega\frac{dV(z)}{dzd\Omega}\int_{M_{\rm lim}}^{\infty}\frac{dn(M,
z)}{dM}dM
\label{eqn:dndz}
\end{equation}
where $dV/dzd\Omega$ is the comoving volume element and $\Delta \Omega$ is the solid angle of the survey.
This $dV/dzd\Omega$ is related to the expansion history of the background cosmology and depends on 
cosmological parameters -- $\Omega_{tot},\Omega_m, h$ and $w(z)$. The halo mass function $\frac{dn}{dM}$ shows 
 exponential sensitivity to the growth of perturbations in the matter density field through $\sigma_8$ and the growth
 function $G(z)$, which again depends on the cosmological parameters. There have been a number of fitting
 forms \cite{Jenkins, Tinker08, Warren, ShethTormen, ShethMoTormen} available in literature obtained from
 both semi-analytic as well as large N-body simulations; in this work we use the fitting form as given in
 reference \cite{Tinker}.

As the mass, $M$, is not directly observed for most of the detected clusters in the sample,
 it is convenient to use a more readily observable proxy for mass, $\mathcal{O}$, such as luminosity,
 temperature or the product of gas mass and temperature in x-rays, $Y_X$ \cite{ChandraVikhlinin}; 
integrated Y parameter in SZE or max S/N in matched filtering of SZ maps \cite{Vanderlinde};
 and richness of red galaxies in optical \cite{Koester, Yee_Ellingson}. Often, the relationship 
between mass and its observable proxy is described a simple power law form as, 
$\log M = A + \alpha \log \mathcal{O} + \gamma \log (1+z)$. Here $A$ is the amplitude, 
$\alpha$ is the slope while the parameter $\gamma$ captures any non-standard (departure from self-similar) 
cluster evolution \cite{MM03}. Since the cluster mass function decreases steeply with mass, a knowledge of 
the distribution of clusters about this relation (or scatter) becomes very important. One frequently uses 
a log-normal model to account for this scatter between mass and its proxy. While in some cases, the values 
of these cluster scaling parameters may be known from past observations to within some error, a slightly 
different choice of the cluster parameters $A, \alpha$ and $\gamma$ from their {\it true} values could 
bring in a significant bias in the derived cosmology. This occurs due to the strong correlation between 
the cluster and cosmology parameters. Thus, these scaling parameters should preferably not be fixed, but 
derived from the data along with the cosmology. However, attempts at marginalizing over these cluster 
parameters to determine the cosmology from only cluster counts, results in a significantly weakened set 
of cosmological constraints. There have been several ideas that have been proposed to break this cluster 
physics -- cosmology degeneracy through self-calibration; for e.g. adding cluster power-spectrum 
 \cite{Feldman_pcl, MM04}, counts-in-cells \cite{LimaHu04}, binning in mass proxy \cite{Hu} as well
 as redshift, planning a survey with both deep and wide component \cite{Satej2010a}, etc. Even a simple mass follow-up of clusters can be very effective at breaking this 
degeneracy and can give a significant improvement in the cosmological constraints \cite{WuRozo}; here, 
we construct a mass follow-up \cite{MM03,MM04}, to calibrate the mass -- observable relation. 
Since both weak lensing as well as x-ray observations can measure the mass like quantity 
$M_f(\theta) = M(\theta)/d_{\rm A}(z)$, where $M(\theta)$ is the halo mass within an angle 
$\theta$, we use this quantity as the observable for constructing our mock catalogs; henceforth we shall
 refer to it as the mass follow-up. We assume a flat 30\% error estimate on $M_f$ in all our mass catalogs 
constructed for each of the surveys considered in section \ref{sec:surveys}. Note, that this is very conservative since with better observations one would be able to achieve more robust estimates of cluster masses \cite{Mahdavi2012a}.

\section{Monte Carlo analysis of mock data}
\label{sec:MCMC}
An accurate and computationally feasible mapping of the underlying
multivariate distribution in a high dimensional parameter space is achieved by constructing
 Markov Chain Monte Carlo (MCMC) chains. A MCMC chain generates a set of points in the
 parameter space which have the same distribution as the target distribution (posterior likelihood).
 Our MCMC chains are obtained by computing the Bayesian likelihood at random points
selected using the Metropolis Hastings (MH) algorithm \cite{MH1, MH2}. For other
 sampling algorithms used in the context of cosmological parameter estimation we refer
 the reader to \cite{Hajian_hmc, Jasche_hmc, Wraith_mcmc,Taylor_hmc,
 Wandelt_gibbs, Jewell_gibbs, Larson_gibbs}. At each point $\Theta =
 (\theta_1, \theta_2,...\theta_n)$ the Bayesian posterior likelihood,  
$\mathcal{L}(\Theta \mid \mathcal{D})$ is computed given the data $\mathcal{D}$ as,
$ \mathcal{L}(\Theta \mid \mathcal{D}) \propto \mathcal{L}(\mathcal{D} \mid
\Theta) \mathcal{L}(\Theta)$. Here ${\mathcal{L}(\Theta)}$ is the prior
likelihood and $\mathcal{L}(\mathcal{D} \mid \Theta)$
 is the likelihood of getting the data $\mathcal{D}$ given the parameter
$\theta$. For number count observations of clusters this is given by the 
Poisson statistic,
\begin{equation}
 \mathcal{L}(\mathcal{D} \mid \Theta) = \prod_m
\frac{{N_m(\Theta)}^{N_m^{(\mathcal{D})}} e^{-N_m(\Theta)}}{N_m^{(\mathcal{D})}}
\label{eqn:dn_like}
\end{equation}
Here, the product is over the redshift bins; $N^{(\mathcal{D})}_m$ corresponds to
the observed number of clusters in the redshift $m$, while $N_m(\Theta)$ is the expected value for a 
model described with the parameter vector $\Theta$. We add to this the chi-sq likelihoods
 $\mathcal{L} = \exp(-\chi^2 /2)$ constructed from the mass follow up data.

The MH algorithm requires a proposal pdf to be specified, which we choose to be a multivariate
 Gaussian distribution for simplicity. For a quicker convergence, the proposal distribution should
 be as close to the actual distribution. In practice, this is achieved by making repeated short runs, each
 time using the computed covariance matrix from the data to be the new proposal matrix. One may also invert the 
Fisher matrix computed for the same observations in order to obtain the proposal covariance matrix,
 see section \ref{sec:Fisher}. The number of iterations required for convergence roughly scales as the number of 
parameters of the model and also depends on the nature of the actual distribution. Longer chains need to be
 constructed when the posterior likelihood departs from a Gaussian distribution. On having sampled sufficiently 
enough points in the parameter space the MCMC chain equilibrates to the target distribution, i.e. 
$\mathcal{L}(\Theta \mid \mathcal{D})$, and is said to have reached convergence. In practice, convergence
 may be tested thorough one of the many convergence tests, see section \ref{sec:discussions}. We refer the 
reader to the references \cite{MCMC1,MCMC2} for further details on the MCMC technique.

\section{Fisher matrix analysis}
\label{sec:Fisher}
The Fisher information matrix \cite{Fisher,Tegmark} is often used to obtain a quick
estimate of the constraints on the parameters of a model for a given experiment. The
Fisher information matrix is defined about a fiducial cosmological model as,
\begin{equation}
{\mathcal{F}}_{ij}\equiv-\Big\langle{\frac{\partial^{2}\ln\mathcal{L}(\Theta \mid
\mathcal{D})}{\partial \theta_{i}\partial \theta_{j}}}\Big\rangle 
\end{equation}
For a large dataset the likelihood function may be approximated as a
 multivariate Gaussian pdf of the model parameters, in which case the covariance matrix of the
model parameters may be constructed as the inverse of the Fisher matrix,
$ {\mathcal{C}}_{ij} \equiv \langle{{{\theta}}_i
\mathcal{\theta}_j}\rangle -
\langle{\mathcal{\theta}_i\rangle\langle\mathcal{\theta}_j}\rangle =
{\mathcal{F}}^{-1}_{ij} $. Using the Cramer-Rao inequality one estimates the marginalized 
standard deviation (s.d.) of the parameters as follows,
$ \Delta \theta \equiv \langle{{{\theta}}^2_i}\rangle -
\langle{{{\theta}}_i}\rangle \langle{{{\theta}}_i}\rangle \geq
{\mathcal{C}}^{1/2}_{ii}$. For galaxy cluster surveys the Fisher 
matrix may be constructed as \cite{Holder01,MM03,MM04}, 
\begin{equation}
{\mathcal F}_{ij} = \Sigma_{m}\frac{\partial N_{m}}{\partial
p_{i}}\frac{\partial N_{m}}{\partial p_{j}}\frac{1}{N_{m}} +
 \Sigma_{k}\frac{\partial M_f^{(k)}}{\partial p_{i}}\frac{\partial M_f^{(k)}}{\partial p_{j}}\frac{1}{\sigma^2_{{M_f}^{(k)}}} + 
\frac{\delta_{ij}}{{\sigma}^2(\theta_i)}
\end{equation}
where $N_{m}$ is the number of observed clusters in each redshift bin $m$; $k$ is the summation index
 over all the follow-up masses; and ${\sigma}(\theta_i)$ is the Gaussian prior on the
 parameter $\theta_i$.

Fisher matrix analysis has the advantage of being computationally simpler as it
is calculated just about a single point, i.e. for the fiducial model parameters.
However it must be emphasized that, this technique provides a reasonably
accurate estimate of parameter constraints only when the likelihood function is
close to being Gaussian. When $3^{rd}$ or higher order moments of the parameters
start becoming important the Fisher analysis may fail. In practice this happens when
the model is described by a large number of degenerate parameters, leading to weakened
constraints due to extended degeneracy between the parameters. However, the likelihood
 distribution of the combination of many independent parameters is always more Gaussian than
 the one of a single parameter \citep[for e.g. ref.][]{Tegmark}. Non-Gaussian likelihoods are also seen when the
 data is not sufficiently large; or for non-Gaussian distribution of errors in the data.
Another disadvantage of Fisher analysis is that it can handle only Gaussian priors;  
many a times, there are parameters which can take values only within a certain range, 
as other values may not be physical, for e.g., in a flat universe, 
$0 \leq \Omega_m \leq 1$. Such flat priors can only be imposed through an MCMC analysis.
In our results we have taken care to see that the comparisons between Fisher and MCMC results are not affected by 
such issues.

\section{Fiducial cosmology, priors and mass follow-up}
\label{sec:fiducial_model}
We adopt our fiducial cosmology from the WMAP 7-year results (Table 6 of
\cite{WMAP7}). For simplicity, we choose a flat Universe
since for an open $w$CDM model, WMAP7+BAO+H0 tightly constraints $\Delta\Omega_{\rm tot}
 \leq 0.007$. However, including the flatness as an additional parameter $\Omega_k$
 with the WMAP prior does not change our results significantly.

We vary the following cosmological parameters - $\Omega_m,\Omega_b, w(z),h,n_s,\sigma_8$ along
 with cluster parameters $A, \alpha, \gamma$. Since cluster counts by themselves do not constrain
 the cosmological parameters $h, \Omega_b$ and $n_s$ very well, we impose 
WMAP Gaussian priors with s.d. 0.013, 0.0016 and 0.013  on these parameters respectively.
 We consider the following cosmological models for making
 comparisons between Fisher and MCMC -- $\Lambda$CDM (dark energy equation of state is
 fixed to be -1), $w$CDM (with a single parameter, $w_0$ dark energy equation of state).
We also consider the dark energy models with a redshift dependent equation of state for which we use the 
two common parametrizations -- the Linder parametrization \cite{CPL} $w(z) = w_0 + w_a \frac{z}{1+z}$
 and the other simple form $w(z) = w_0 + w_z z$. In the rest of the paper, for the $w$CDM 
model we shall always use $w_0$ even to denote the single parameter dark energy equation of
 state, to ensure notational consistency. 

 We model a log-normal scatter \cite{LimaHu05} in the mass-observable relations parametrized as \cite{Sartoris_ng}, 
$\eta(z) = \eta_0(1+\eta_1 z)$. Due to the steep form of the cluster mass function, $\frac{dn}{dM}$ cluster 
counts are very sensitive to the amount of scatter. Thus when analysing the cluster count data the scatter
 should preferably not be fixed, but determined jointly with the cosmology. This is especially true for 
 cluster data in optical/IR where the scatter is seen to be large \cite{Rozo10,Gladders}. However, this scatter
 need not be assumed to be completely unknown; it is expected that the mass follow-up observations would 
yield some constraints on the amount of scatter in the $M- \mathcal{O}$ relation. Thus, we place Gaussian 
priors with s.d. of 0.1 and 0.12 on the log-normal scatter of 0.45 and 0.58 in DES and RCS2 respectively.
In our analysis of mock cluster data from the x-ray and SZE surveys, the relatively small scatter $\eta$ is kept fixed.

 For computing the $\frac{dN}{dz}$ likelihoods we first generate a mock dataset cluster distribution in redshift bins of
 width 0.1. We then use eq. \ref{eqn:dn_like} to compute the likelihoods for each point $\Theta$ in the 
 Markov chain. As mentioned in section \ref{sec:dndz}, just cluster counts cannot place tight constraints on cosmology by
 themselves. To break the cosmology-cluster physics degeneracy we consider a simple mass follow-up, $M_f$, of
 100 clusters for all our surveys except the WFXT. For the WFXT we consider a larger follow-up of 1,000 clusters.
 Our mass catalogs are uniformly sampled over mass (from $2-8 \times 10^{14} ~{\rm M}_{\odot}$) and redshift
 ($0.3 \leq z \leq 0.9$) and consists of a table of redshift, mass, and observable mass-proxy. The likelihood of 
  $M_f$  is then added to the likelihoods from $\frac{dN}{dz}$.

\section{Cluster surveys}
\label{sec:surveys}

\subsection{Large Yield Surveys}

We examine in detail, on a case by case basis, the model dependent cosmological parameter constraints
 from various surveys. We focus mainly on upcoming surveys, especially in x-ray and optical, which promise to provide survey yields of
 ten-to-hundreds of thousands of clusters of galaxies. These are the surveys which will have enough number of clusters so as to
 self-calibrate multiple cluster specific nuisance parameters to give {\it unbiased} cosmological parameter constraints.
 We consider two survey cases each for x-ray and optical
based observations. In x-ray we consider the very large surveys, the upcoming eROSITA and the proposed WFXT. For optical surveys we
 consider the RCS2 survey whose observation runs have been recently completed and the DES survey that will yield clusters in the near future . 

  Table \ref{tab:surveys} summarizes the details like area coverage, expected
 cluster detections, redshift range, etc. that were assumed for each of the cluster surveys investigated in this paper.

 \begin{table}[ht]
\hfil
\begin{tabular}{|c|c|c|c|c|c|c|}
\hline 
Survey & wavelength & ${\rm N_{cl}}$  & area & flux/${\rm B_{gc}}$/${\rm N_{200}}$ & redshift & ${\rm N_{follow-up}}$ \\
  & & (in k) & (${\rm deg.}^2$) & cut & range & clusters \\
\hline
\hline 
eROSITA & x-ray   		& 120	& 27,000	& $4\times10^{-14}$ erg-cm$^{\text{-2}}$-s$^{\text{-1}}$ & 0.1 - 1.3 &	100\\
WFXT 	& x-ray	  		& 300	& 20,000	& $5\times10^{-15}$ erg-cm$^{\text{-2}}$-s$^{\text{-1}}$ & 0.1 - 1.3 & 1,000\\
RCS2 	& optical 		& 15	& 1,000		& $B_{gc}=300$	& 0.1 - 1.0 & 100\\
DES 	& optical 		& 155	& 5,000		& $N_{200}=17$	& 0.1 - 1.3 & 100\\
Planck 	& SZE 	& 2		& 32,000	& 300 mJy	& 0.1 - 1.3 & 100\\
\hline
\end{tabular}
\hfil
\caption{A summary of the six cluster surveys that we consider in this work along with the relevant parameters.}
\label{tab:surveys}
\end{table}

\subsubsection{X-ray surveys}
Clusters are detected in x-ray due to the thermal Bremsstrahlung emission from the hot ICM. X-ray telescopes
 like ROSAT, XMM-Newton and Chandra along with others have observed a large number of clusters at various 
redshifts. Some of these observations have already given interesting independent constraints on cosmological
 models \cite{ROSAT_cosmo, Chandra_cosmo}.

For a given survey the limiting mass $M_{\rm{lim}}(z)$ in eq. \ref{eqn:dndz} is found from the 
flux limit, $f_{\rm lim}$ of the survey. For x-ray survey, we adopt 
luminosity-mass relations from \cite{ChandraVikhlinin} given by the following expression\footnote{An alternate
 way is to take $\gamma$ directly as the exponent of $E(z)$ without the
 need for the $(1+z)^{\gamma}$ term; however we avoid this form as we
 find that it induces extra correlations of $\gamma_X$ with the cosmological
 parameters - $\Omega_m$, $h$ and others.}
with fiducial parameter values: $A = -4.24$, $\alpha = 1.61$ and 
$\gamma = 0$ with a log-normal scatter of 0.246.
\begin{equation}
L_X=10^{A} \left( \frac{M_{500}}{1 \times 10^{15}} \right )^{\alpha}E^{1.85}
(z) (1+z)^{\gamma} 
\end{equation}

\paragraph{eROSITA :}
 The extended Roentgen Survey with an Imaging Telescope Array (eROSITA) is
the next big full sky x-ray survey, expected to be launched in the near future \cite{Erosita2012}. The primary
 scientific goal would be to study the nature of dark energy using
 about 100,000 x-ray galaxy clusters. We model the survey with a limiting flux
of $f_{\rm lim}=4\times10^{-14}$ erg cm$^{\text{-2}}$ s$^{\text{-1}}$
 in the [0.5-2.0 keV] band with a sky coverage of $\sim$ 27,000 deg$^{\text{2}}$.

\paragraph{WFXT :}
 The Wide Field X-ray Telescope (WFXT) is a proposed x-ray mission which is
expected to be 2 orders of magnitude more sensitive than any previous x-ray mission \cite{WFXT_cosmo}. It would
survey a large fraction of the sky at a high angular resolution with a deep, wide and a
medium survey. The proposed survey would be in 3 parts -- a deep survey of 100 sq. deg, 
a medium survey of 3,000 sq. deg and a wide survey of 20,000 sq. deg. 
We examine the constraints from the wide survey with a flux limit of 
 $f_{\rm lim}=5\times10^{-15}$ erg cm$^{\text{-2}}$ s$^{\text{-1}}$ in the
[0.5-2.0 keV] band.

In both the x-ray surveys we impose an additional flux cut-off corresponding to
a mass limit of  ${8\times 10^{13}  h^{-1}}~ {\rm M_{\odot}}$. This is to prevent selection of very low mass halos
like groups of galaxies, especially at lower redshifts.

\subsubsection{Optical surveys}

Clusters are detected in optical surveys by searching for over-densities of
galaxies with known color properties. Compared to other wavelengths, 
cluster surveys in optical provide the advantage of a high S/N ratio and a
 wide field of view. Also, because of a greater depth, these surveys give the largest
 yield in terms of cluster detection per sq. deg. Optical surveys are also essential to provide
 photometric redshift information of clusters detected in other wavelengths and for the measurement
 of galaxy power spectrum.

\paragraph{RCS2 :}
The Red Sequence Cluster Survey 2 (RCS2) is the sequel to RCS1 which was a 78 sq. deg. survey which
 detected $\sim$ 1,000 clusters. The RCS2 is a larger survey with 1,000 sq. deg of sky coverage carried out
 using the one-square-degree MegaCam on the CFHT. As with the RCS1, it uses a red-sequence of cluster early-type
 galaxies to identify galaxy clusters \cite{RCS2survey}. It is expected to detect 15,000 clusters in the redshift range 0.3 - 1.0. We model the RCS2 on the lines of RCS1 (see \cite{Gladders}) \cite{Yee_Ellingson} with a uniform $B_{gcR} = 300$
 cut-off as follows, with $A=10.29$, $\alpha=1.70$, $\gamma=0.64$ and $\eta_0 = \sigma_{M_{200}|Bgc} = 0.58$.
\begin{equation}
 M_{200} = 10^{A} B_{gcR}^{\alpha} (1+z)^{\gamma}
\end{equation}

\paragraph{DES :}
The Dark Energy Survey (DES) is an upcoming optical imaging survey project that would
map out 5,000 sq. deg. of area in 5 years using the Blanco 4-meter telescope
 at the Cerro Tololo Inter-American Observatory in Andes. One of the primary
 goal of this survey is to make precision measurements of dark energy using
 various cosmological probes like Supernovae, baryon acoustic oscillations,
 weak lensing and galaxy clusters \cite{DES_cosmo}. We model this survey with a maxBCG selection
 function and a mass cut-off corresponding to $N_{200}=17$ as used in ref. \cite{Rozo07},
with $A=2.34$ and $\alpha=0.757$ and $\gamma=0$ and a log-normal scatter,  
$\eta_0 = \sigma_{M_{200}|N_{200}} = 0.45$ in the mass-richness relation. 
Larger datasets give the opportunity to play with greater number of nuisance parameters for more robust cosmological constraint forecasts. Thus, 
for the DES we also examine
 the effect of introducing an extra nuisance parameter $\eta_1$ (as described in section \ref{sec:fiducial_model}) that can capture
 the redshift dependent scatter.
 
\begin{equation}
 \ln(N_{200}) = A + \alpha \ln \left( \frac{M_{200}}{1.09 \times 10^{14}} \right) + \gamma \ln(1+z) 
\end{equation}

\subsection{Smaller Yield SZE Surveys}

For the sake of completeness and to explore survey complementarity, we also consider the ongoing space based Planck mission. The Planck survey has already detected
clusters \cite{Planck_clusters} through their SZE signal and have obtained initial mass calibration \cite{Planck_masscal}. 
 For the Planck survey, we use the integrated Compton $y$ parameter -- $Y$ as
 the mass proxy and write the SZE flux-mass relation as in \cite{MM03,MM04},
\begin{equation}
 Y d_{\rm A}^{2}=f(\nu) 10^{A}  M_{200}^{\alpha}E^{2/3}(z)\left(1+z\right)^{\gamma}
\end{equation}

Here, $f(\nu)$ is the well known frequency dependence of the SZ effect and $d_{\rm A}$ is 
the angular diameter distance in Mpc. The conversion from Y to SZ flux is done as in ref. \cite{WhiteMajumdar}. 
The values for the SZE scaling parameters are: $A = -28.07$, $\alpha = 1.61$, $\gamma = 0$ 
which are taken from recent observations \cite{anya}. A log-normal scatter
 of 0.2 is assumed in the flux-mass relation.



\paragraph{Planck :}

Planck covers the entire sky, with good sensitivity and
 at a high angular resolution over 9 frequency channels from 
30--857 GHz. Planck was launched into orbit in 2009, and is expected to
 complete its observations by 2012. Among other goals such as measurements of intensity and
polarization of the primordial and lensed CMB, Planck would also create a
catalog of galaxy clusters through SZE. The first all sky SZ catalog from Plank was released recently which was from six months of survey.
Projecting the initial cluster count to the full Planck mission time, one would expect a total yield of $1000-2000$ clusters over the entire sky. We take the flux limit  
to be 160 mJy (at 353 GHz) which gives $\lesssim$ 2,000 clusters \cite{Chamballu} in 
$\sim$ 32,000 deg$^{\text{2}}$. The higher flux limit means that Planck
 would be able to detect only the most massive clusters.

\section{Results}
\label{sec:results}
We now compare the forecasts for parameter constraints from each of the surveys considered in section \ref{sec:surveys}, 
 obtained from Fisher and MCMC analysis of mock cluster data.

 \begin{table}[ht]
\hfil
\begin{tabular}{|c|cc|cc|cc|}
\hline model & \multicolumn{2}{c|}{$\Lambda$CDM} & \multicolumn{2}{c|}{$w$CDM} & \multicolumn{2}{c|}{$w_a$CDM} \\
parameter & MCMC & Fisher & MCMC & Fisher & MCMC & Fisher \\
\hline
\hline
$\Omega_m$ 	& 0.028 	& 0.026	& 0.035	& 0.028	& {\bf 0.080}	& {\bf 0.028}	\\
$w_0$ 		& ---		& ---	& 0.191	& 0.145	& {\bf 0.374}	& {\bf 0.172}	\\
$w_a$ 		& --- 		& --- 	& ---	& ---	& {\bf 0.685}	& {\bf 0.165}	\\
$\sigma_8$ 	& 0.016 	& 0.018	& 0.034	& 0.025	& {\bf 0.059}	& {\bf 0.026}	\\
$A$ 		& 0.048 	& 0.036	& 0.052	& 0.039	& 0.096 	& 0.055 \\
$\alpha$ 	& 0.031 	& 0.035	& 0.037	& 0.040	& 0.066		& 0.047	\\
$\gamma$ 	& 0.100		& 0.089	& 0.179	& 0.177	& 0.204		& 0.185	\\
\hline
\end{tabular}
\hfil
\caption{A comparison of the standard deviations of marginalized parameter constraints from analysis of the 
mock cluster data from the eROSITA x-ray survey using MCMC and Fisher methods. Numbers in bold highlight the
 significant discrepancy in constraint estimates between the two methods.}
\label{tab:eROSITA_results}
\end{table}

\begin{figure}[ht]
\centering
  \includegraphics[width=15cm]{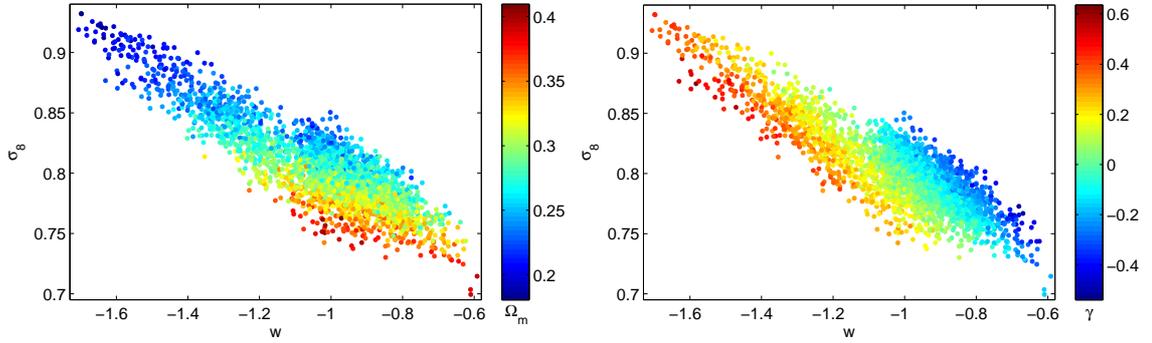}
\caption[]{The degeneracy between the cosmological parameters $\Omega_m$ and $\sigma_8$ for the eROSITA 
survey is indicated by a few samples drawn from the posterior likelihood distribution of the $w$CDM model.
 The correlation with a third parameter -- $\Omega_m$ (left) and $\gamma$ (right) is indicated through the 
color or value of the third parameter.}
\label{fig:eROSITA_3D}
\end{figure}

\begin{figure}[ht]
\centering
  \includegraphics[width=15cm]{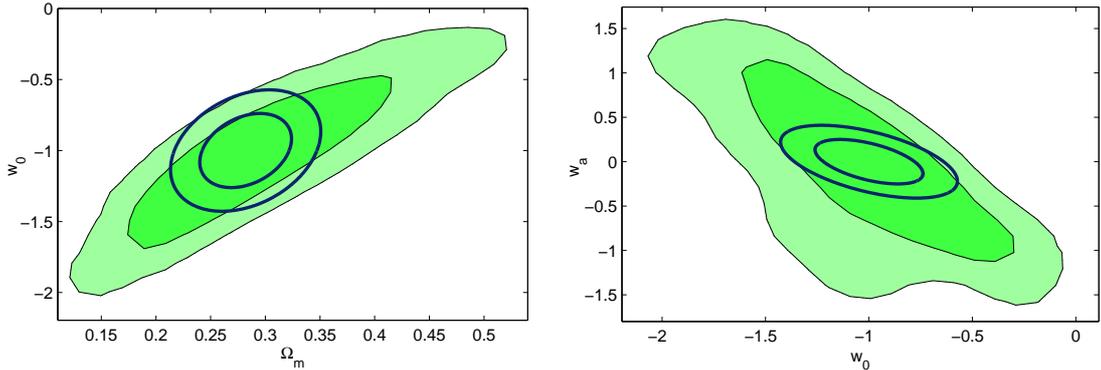}
\caption[]{The dark(light) green regions indicate the projected 1-$\sigma$ (2-$\sigma$) confidence regions obtained from
 the analysis of mock cluster data for the eROSITA survey using MCMC methods for a $w_a$CDM model. The blue ellipses are
 the corresponding constraints from a Fisher analysis.}
\label{fig:eROSITA_discrep}
\end{figure}

\begin{figure}[ht]
\centering
  \includegraphics[width=15cm]{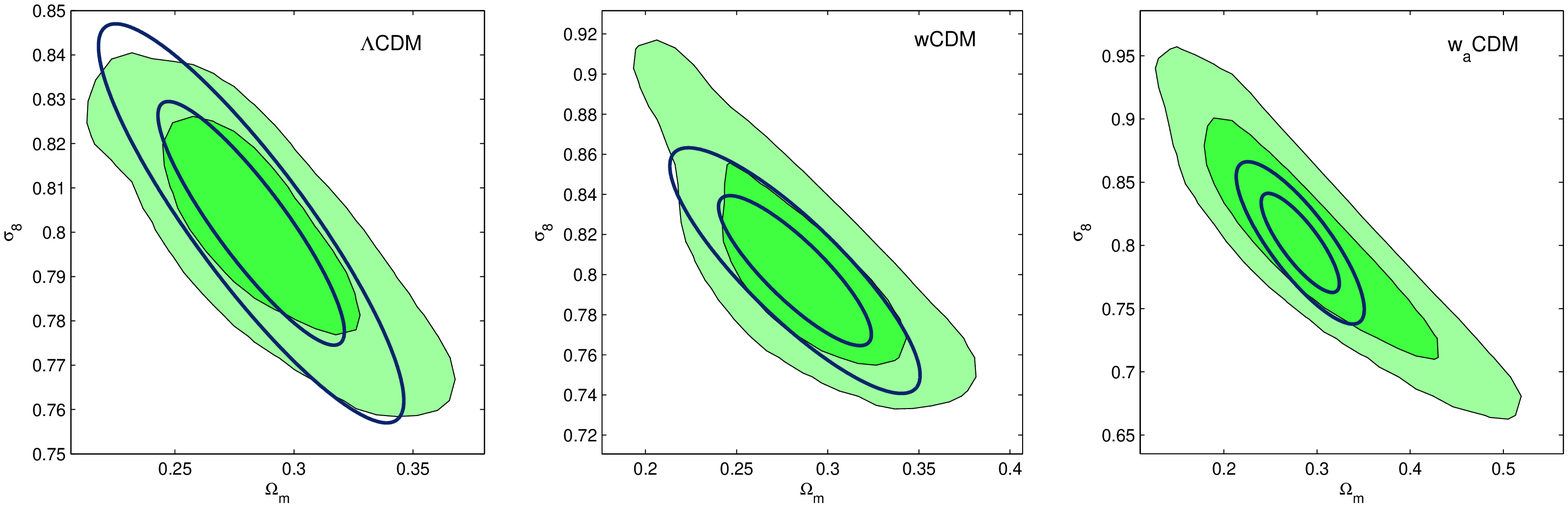}
\caption[]{The dark (light) green regions indicate the projected 1-$\sigma$ (2-$\sigma$) confidence regions obtained
 from the analysis of mock data for eROSITA using MCMC methods for the three cosmological models -- $\Lambda$CDM,
 $w$CDM and $w_a$CDM in the $\Omega_m-\sigma_8$ plane. The blue ellipses are the corresponding constraints from a 
Fisher analysis.}
\label{fig:eROSITA_models}
\end{figure}

\subsection{X-ray surveys}
\paragraph{eROSITA :}
The eROSITA survey places strong constraints on the $\Lambda$CDM cosmology, with $\Delta \Omega_m$
 = 0.028 and $\Delta \sigma_8$ = 0.016. To compare, these constraints are slightly 
better than those obtained from just the CMB (WMAP7 results) for which $\Delta \Omega_m$ =
 0.029 and $\Delta \sigma_8$ = 0.030. 
For the $w$CDM model, the constraints get relaxed to 0.035 and 
0.034 on $\Omega_m$ and $\sigma_8$ respectively, with almost a twofold increase in $\Delta \sigma_8$. This 
occurs due to the long degeneracy between $w_0$ and $\sigma_8$, that causes wider constraints on $\sigma_8$ 
after marginalization over $w$; the marginalised constraint on $w_0$ is 0.191. This degeneracy in the $w_0 - \sigma_8$ plane
 is indicated through figure \ref{fig:eROSITA_3D}; the correlation of $\sigma_8$ and $w_0$ with other parameters
 like $\Omega_m$ and $\gamma$ is also shown here. The expected degeneracy of $\Omega_m - \sigma_8$ and 
$w - \gamma$ is clearly visible through the color coding. We find here that the Fisher results are 
underestimated by a factor of $\approx$ 1.3 for all $\Omega_m$, $\sigma_8$ and $w_0$. Addition of priors 
from the WMAP7 results can significantly shrink the constraints on the $w$CDM model to $\Delta \Omega_m$ 
= 0.013, $\Delta w_0$ = 0.042 and $\Delta \sigma_8$ = 0.011. 
The large number of clusters expected to be detected from a full sky survey like eROSITA will also
 place constraints on the possible evolution of the dark energy equation of state. We examine the constraints
from $w_a$CDM model in which dark energy equation of state varies smoothly from $w_0$ at $z=0$ to approach $w_0+w_a$
 at a large redshift ($z \gg 1$).
For the $w_a$CDM model, with a two parameter dark energy
 equation of state there would be weaker constraints on cosmology. We also see that there is no 
agreement between Fisher and MCMC estimates whatsoever -- $\Delta \Omega_m$ = 0.080(0.028), $\Delta w_0$ = 
0.374(0.172), $\Delta w_a$ = 0.685(0.165) and $\Delta \sigma_8$ = 0.059(0.026) for the MCMC(Fisher) results 
respectively. There is also a noticeable difference in the degeneracy direction in the $w_0 - w_a$ plane, see 
figure \ref{fig:eROSITA_discrep}. It is alarming to note how {\it the Fisher estimates mislead us into false 
estimates of tight constraints on the parameter $w_a$, where the two forecasts can differ by a factor up to 4 or 
 more}. We find fairly good agreement between the Fisher and MCMC results only for the $\Lambda$CDM model. The 
results between the two methods are seen to progressively diverge for the $w$CDM and $w_a$CDM models, see 
figure \ref{fig:eROSITA_models}. The detailed results for eROSITA are also listed in table \ref{tab:eROSITA_results}.

\begin{table}[ht]
\hfil
\begin{tabular}{|c|cc|cc|cc|cc|}
\hline model & \multicolumn{2}{c|}{$\Lambda$CDM} & \multicolumn{2}{c|}{$w$CDM} & \multicolumn{2}{c|}{$w_a$CDM} & \multicolumn{2}{c|}{$w_z$CDM}  \\
parameter & MCMC & Fisher & MCMC & Fisher & MCMC & Fisher & MCMC & Fisher \\
\hline
\hline
$\Omega_m$ 	& 0.012 & 0.017	& 0.029	& 0.026	& {\bf 0.047} 	& {\bf 0.026}	& 0.034	& 0.043	\\
$w_0$ 		& ---	& ---	& 0.133	& 0.120	& {\bf 0.225}	& {\bf 0.120}	& 0.181	& 0.206	\\
$w_a$ 		& --- 	& --- 	& ---	& ---	& {\bf 0.343}	& {\bf 0.093}	& 0.172	& 0.125	\\
$\sigma_8$ 	& 0.014	& 0.017	& 0.031	& 0.029	& {\bf 0.041}	& {\bf 0.029}	& 0.037	& 0.043	\\
$A$ 		& 0.024 & 0.028	& 0.039	& 0.035	& 0.052 	& 0.037		& 0.045	& 0.053	\\
$\alpha$ 	& 0.037 & 0.037	& 0.038	& 0.038	& 0.039		& 0.038		& 0.039	& 0.039	\\
$\gamma$ 	& 0.048 & 0.057	& 0.136	& 0.137	& 0.140		& 0.153		& 0.145	& 0.154	\\
\hline
\end{tabular}
\hfil
\caption{A comparison of standard deviations of marginalized parameter constraints obtained using MCMC and Fisher
analysis of the mock cluster data from the WFXT x-ray survey. Numbers in bold highlight the 
significant discrepancy in constraint estimates.}
\label{tab:WFXT_results}
\end{table}

\begin{figure}[ht]
  \centering
{
    \subfigure[]{
    \centering
    \includegraphics[width=7.3cm]{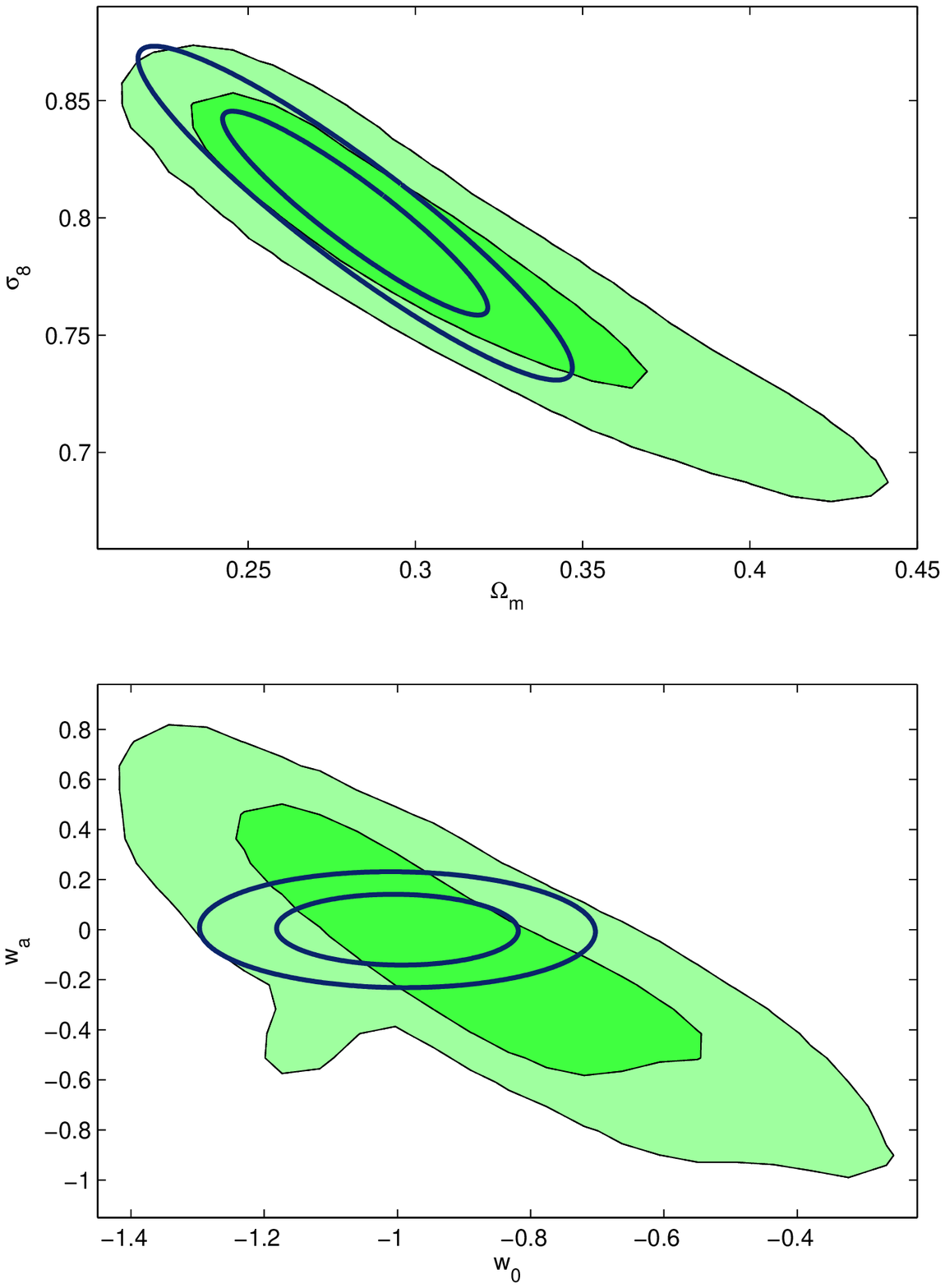}
    }
    \subfigure[]{
     \centering
      \includegraphics[width=7.3cm]{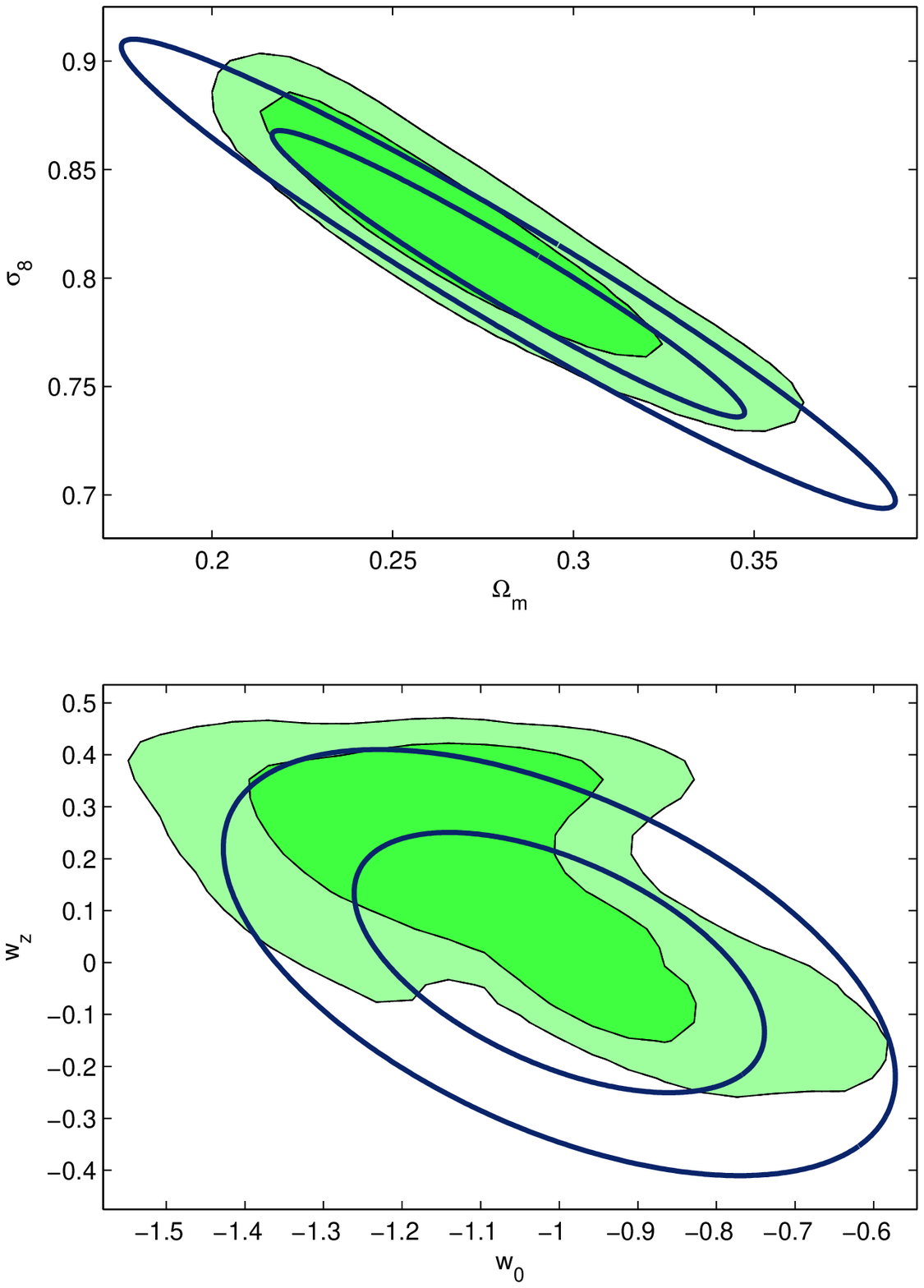}
    }
\caption[]{A comparison of parameter constraints from the WFXT survey for the two different parametrizations
 of the dark energy equation of state -- $w(z)=w_0 + \frac{w_az}{1+z}$ (left) and $w(z)=w_0 + w_z z$ (right). 
The dark(light) green regions indicate the projected 1-$\sigma$ (2-$\sigma$) confidence regions obtained from the 
analysis of mock cluster data using MCMC methods. The blue ellipses indicate the corresponding constraints from a Fisher 
analysis.}
\label{fig:WFXT_DE_EOS}
}
\end{figure}

\begin{figure}[t]
\vspace*{-1.2cm}
\centering
\hspace*{-2.1cm}
  \includegraphics[width=19cm]{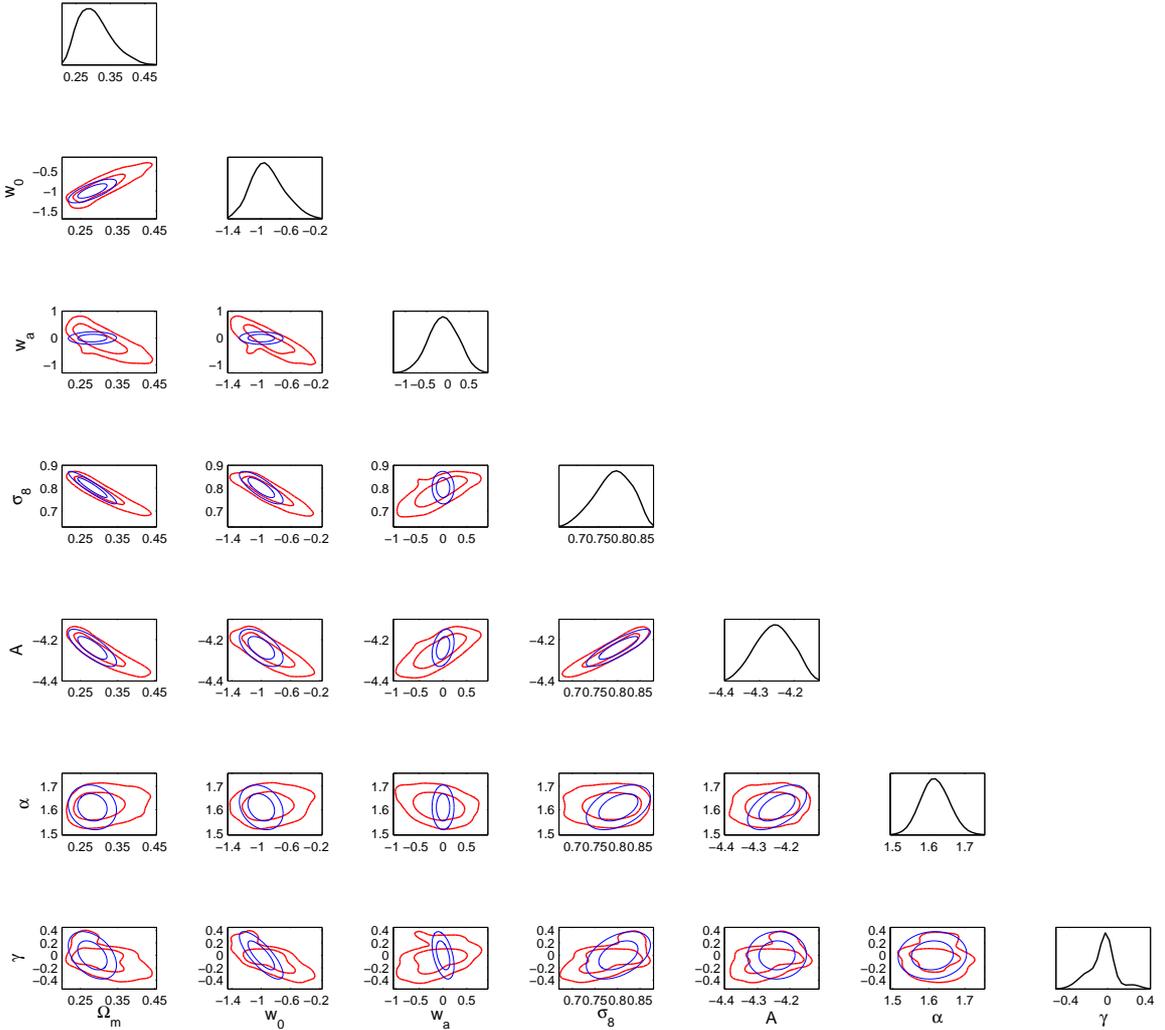}
\vspace*{-2.2cm}
\caption[]{Projected 1-$\sigma$ and 2-$\sigma$ confidence regions obtained from the analysis of mock cluster data 
for the WFXT survey for the $w_a$CDM model. The contours in red (blue) show results obtained using MCMC (Fisher) 
analysis. The diagonal plots in black show the corresponding marginalized pdf's for each parameter.}
\label{fig:triplot_WFXT_DE_}
\end{figure}

\paragraph{WFXT :}
The WFXT is seen to place much stronger constraints on cosmology -- i) By 
virtue of a large number of cluster detections due to its high flux sensitivity; ii) a 10 times 
larger mass follow-up of 1,000 clusters that we have considered to break the cosmology - cluster physics degeneracy. 
On the $\Lambda$CDM model we get very tight constraints with $\Delta \Omega_m$ = 0.012 and
 $\Delta \sigma_8$ = 0.027. These constraints are about a factor of 2 or so better as compared to eRSOITA. 
Here, the Fisher case slightly overestimates for the constraints as compared
 to MCMC. For the $w$CDM model we find $\Delta \Omega_m$ = 0.029, $\Delta \sigma_8$ = 0.031 and
 $\Delta w_0$ = 0.133; while the Fisher and MCMC give very similar results. However
 they turn out to be very different for the $w_a$CDM model, as seen in the eROSITA case before. Here, we 
find $\Delta \Omega_m$ = 0.047(0.026), $\Delta w_0$ = 0.225(0.120),  $\Delta w_a$ = 0.343(0.093) 
 and $\Delta \sigma_8$ = 0.041(0.029) for the MCMC(Fisher) forecasts respectively. The only difference 
is that the estimates for $\sigma_8$ are not as divergent. We find again that introducing the parameter
 $w_a$ for the redshift evolution of dark energy equation of state makes the Fisher constraints
 seem much tighter than they actually are; also we find that the {\it degeneracy direction predicted 
by the Fisher method is incorrect}, see figure \ref{fig:WFXT_DE_EOS}. 
Figure \ref{fig:triplot_WFXT_DE_} 
shows the 2-dimensional likelihood contours for pairs of some of the relevant parameters, along 
with the 1-dimensional marginalized pdf's along the diagonal. In the figure, for the 2-dimensional likelihoods, 
 the red and blue contours indicate the 1-$\sigma$ and 2-$\sigma$ confidence regions for MCMC and 
Fisher forecasts respectively.

For the WFXT survey we also examine the effect of a different parametrization for the dark energy equation of state: $w(z) = w_0 + w_z z$.
We do so since both models, $w_a$ and $w_z$ have been considered in the literature, and here we point
 out the degeneracies resulting from the two parametrizations. The $w_z$ parametrization goes linearly with $z$,
 clearly not realistic at very large $z$ ($z \gg 1$). However as long as we are dealing with low redshift data,
 and we are interested in constraining deviations from a $w=-1$ equation of state, both models can be used.

\alert{The degeneracy direction of Fisher ellipses in the $w_0$-$w_a$ plane is  different from that in the $w_0$-$w_z$ plane.}
 \alert{Although the Fisher ellipses do not provide a very accurate description for the dark energy constraints for both the parametrizations (see Fig. \ref{fig:WFXT_DE_EOS}),}
we find that the $w_z$ parametrization compares much favorably as far as the discrepancy between Fisher/MCMC forecasts from cluster
 surveys are considered. For e.g. we find that $\Delta \Omega_m$ = 0.034(0.043), $\Delta w_0$ = 0.225(0.140),  $\Delta w_z$ = 0.343(0.075) 
 and $\Delta \sigma_8$ = 0.041(0.031) for the MCMC(Fisher) forecasts respectively.  For the $w_z$CDM model
  not only is the discrepancy significantly smaller, but also the degeneracy directions are more or 
less correctly predicted by the Fisher method, see figure \ref{fig:WFXT_DE_EOS}. However, the marginalized 
2-dimensional likelihoods in the dark energy plane are seen to be highly skewed.
Table \ref{tab:WFXT_results} summarises the results for the WXFT survey. 
\alert{Finally, we emphasize again that the computation of the Fisher matrix (and therefore the constraints resulting from it)
are dependent on the derivatives computed just at the Fiducuial point in the parameter space. For a non-Gaussian
distribution of the likelihood, the Fisher matrix may or may not provide a correct estimate of the actual (MCMC) constraints.
This may be the reason for the different discrepancies in the Fisher/MCMC comparison for the two parametrizations, as seen here.
See also the discussion under DES in section \ref{sec:opticalresults}.}

\begin{table}[ht]
\hfil
\begin{tabular}{|c|cc|cc|}
\hline model & \multicolumn{2}{c|}{$\Lambda$CDM} & \multicolumn{2}{c|}{$w$CDM} \\
parameter & MCMC & Fisher & MCMC & Fisher \\
\hline
\hline
$\Omega_m$ 	& 0.040 & 0.040 & 0.046	& 0.040	\\
$w_0$ 		& ---	& ---	& 0.144	& 0.137	\\
$\sigma_8$ 	& 0.141 & 0.140 & 0.153	& 0.151	\\
$A$ 		& 0.299 & 0.300 & 0.302	& 0.301	\\
$\alpha$ 	& 0.117 & 0.117 & 0.117	& 0.117	\\
$\gamma$ 	& 0.210 & 0.220 & 0.227	& 0.221 \\
$\eta_0$ 	& 0.099 & 0.114 & 0.111	& 0.120	\\
\hline
\end{tabular}
\hfil
\caption{A comparison of standard deviations of marginalized parameter constraints obtained using MCMC and Fisher
analysis of the mock cluster data from the RCS2 optical survey for the $w$CDM model.}
\label{tab:RCS2_results}
\end{table}

\begin{figure}[t]
\vspace*{-1.2cm}
\centering
  \hspace*{-2.1cm}
  \includegraphics[width=19cm]{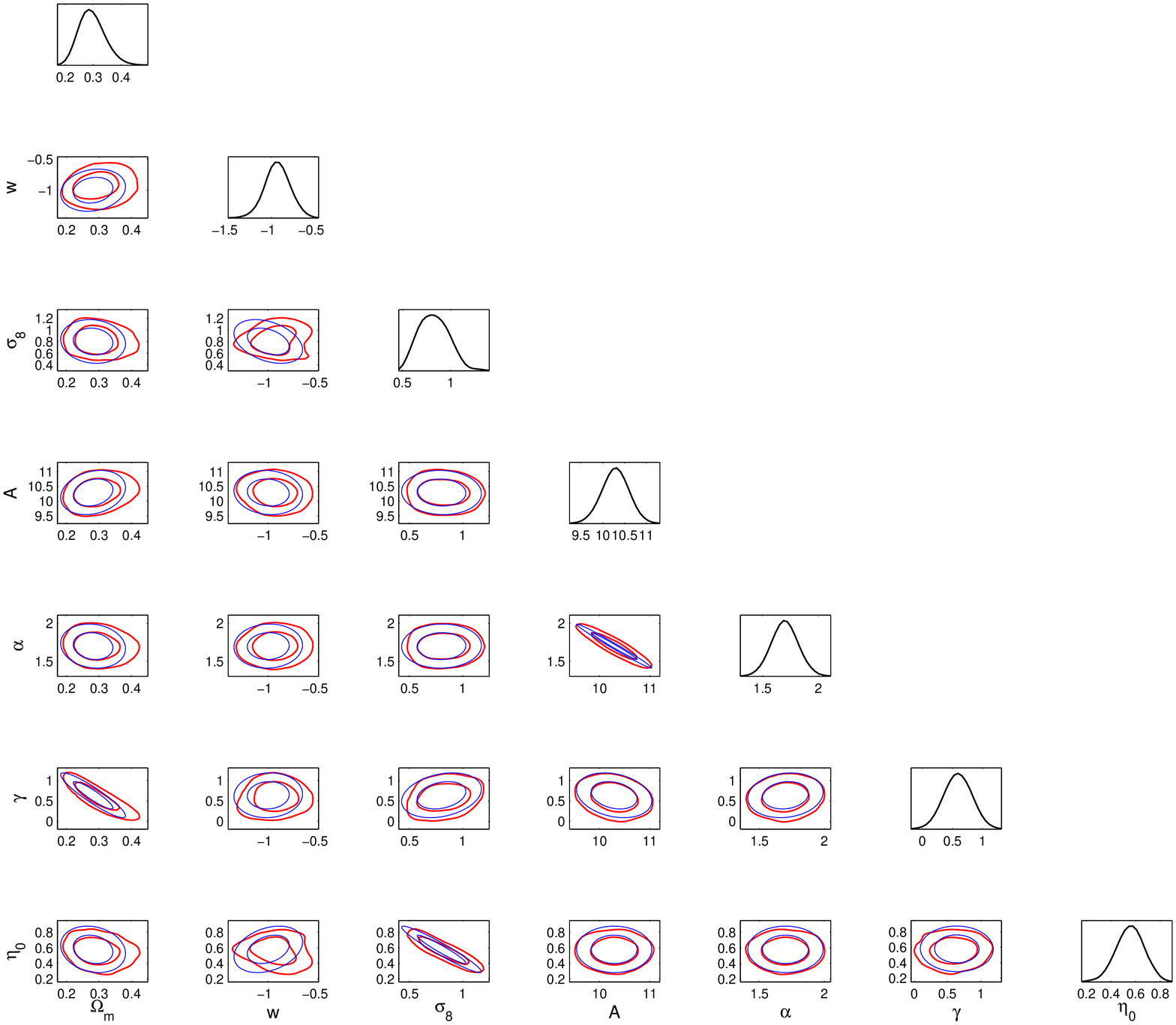}
\vspace*{-2.2cm}
\caption[]{Projected 1-$\sigma$ and 2-$\sigma$ confidence regions obtained from the analysis of mock data
 for the RCS2 survey for the $w$CDM model. The contours in red (blue) show results obtained using MCMC 
(Fisher) analysis. The diagonal plots in black show the corresponding marginalized pdf's for each parameter.}
\label{fig:triplot_RCS2}
\end{figure}

\subsection{Optical surveys}
\label{sec:opticalresults}
For the optical surveys there is a large scatter between the mass-observable relation. To account for this we
 introduce two additional parameters to characterize this scatter and its evolution with redshift. As we shall see,
 the presence of an unknown scatter creates extra degeneracies, and weakens the cosmological constraints. In order to
investigate this degeneracy, we consider here only the $\Lambda$CDM and $w$CDM model for simplicity.
\paragraph{RCS2 :}
For the RCS2 we see a reasonably good agreement between the Fisher and MCMC forecasts. For the $\Lambda$CDM model we
 find fairly wide constraints of 0.040 and 0.141 on $\Omega_m$ and $\sigma_8$ respectively, which get marginally relaxed
 to 0.046 and 0.153 in the $w$CDM model with $\Delta w_0$ being 0.144. It is interesting to note that the
 constraints on $\sigma_8$ are much wider than those from smaller yield SZ surveys like  Planck. This is
 caused due to presence of large scatter in the mass - proxy relation which causes a large degeneracy between 
 $\sigma_8$ and the scatter $\eta_0$; marginalizing over this degeneracy washes out the constraints on $\sigma_8$. Figure 
\ref{fig:triplot_RCS2} shows the 2-dimensional likelihood contours for each pair of some of the relevant parameters, along 
with the 1-dimensional marginalized pdf's along the diagonal. For the 2-dimensional likelihoods, the red and blue contours indicate 
the 1-$\sigma$ and 2-$\sigma$ confidence regions for MCMC and Fisher forecasts respectively. The RCS2 results are listed in table 
\ref{tab:RCS2_results}.

\begin{table}[ht]
\hfil
\begin{tabular}{|c|cc|cc|cc|cc|}
\hline model & \multicolumn{2}{c|}{$\Lambda$CDM} & \multicolumn{2}{c|}{$\Lambda$CDM} & \multicolumn{2}{c|}{$w$CDM} & \multicolumn{2}{c|}{$w$CDM} \\
parameter & MCMC & Fisher & MCMC & Fisher & MCMC & Fisher & MCMC & Fisher \\
\hline
\hline
$\Omega_m$ 		& 0.025 	& 0.028		& 0.030 & 0.032 & 0.033 	& 0.035 	& 0.037 & 0.036	\\
$w_0$ 			& --- 		& --- 		& ---	& --- 	& {\bf 0.071} 	& {\bf 0.051} 	& 0.149 & 0.130	\\
$\sigma_8$ 		& {\bf 0.070} 	& {\bf 0.101}	& 0.113 & 0.118 & 0.094 	& 0.109 	& 0.116 & 0.123	\\
$A$ 	& 0.082 	& 0.080		& 0.095 & 0.099 & 0.102 	& 0.100 	& 0.104 & 0.100	\\
$\alpha$ 		& 0.055 	& 0.050		& 0.055 & 0.052 & 0.055 	& 0.052 	& 0.056 & 0.052	\\
$\gamma$ 		& 0.099 	& 0.124		& 0.131 & 0.146 & 0.140 	& 0.143 	& 0.148 & 0.146	\\
$\eta_0$ 		& 0.057 	& 0.084		& 0.091 & 0.100 & 0.075 	& 0.097 	& 0.089 & 0.100	\\
$\eta_1$ 		& --- 		& --- 		& 0.041 & 0.026 & ---		& --- 		& 0.072 & 0.066 \\
\hline
\end{tabular}
\hfil
\caption{Comparison of $1-\sigma$ marginalized parameter constraints from MCMC and Fisher
analysis of the mock cluster data from the DES optical survey. Numbers in bold highlight 
the significant discrepancy in forecasts.}
\label{tab:DES_results}
\end{table}

\begin{figure}[ht]
\centering
  \includegraphics[width=15cm]{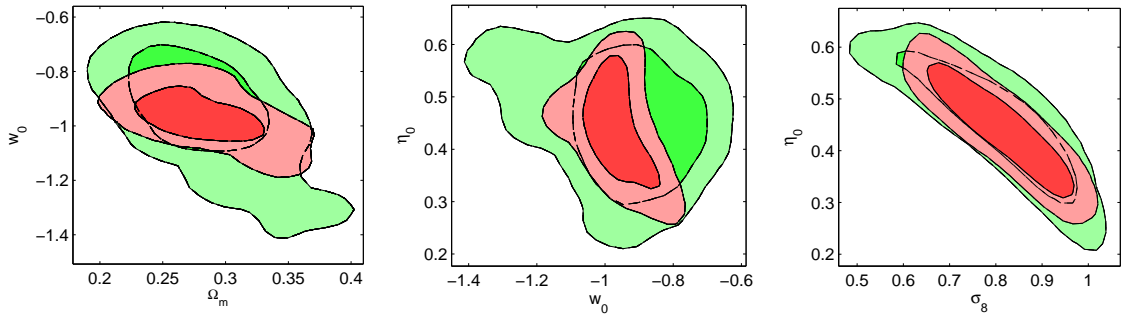}
\caption[]{The dark (light) regions indicate the projected 1-$\sigma$ (2-$\sigma$) confidence 
regions obtained from the analysis of mock data for the DES survey using MCMC methods for the 
$w$CDM model in various parameter planes. The green(red) region indicates constraints with 
(without) redshift dependent scatter. 
}
\label{fig:DES_eta1}
\end{figure}

\begin{figure}[ht]
  \centering
{
    \subfigure[]{
    \centering
    \includegraphics[width=7cm]{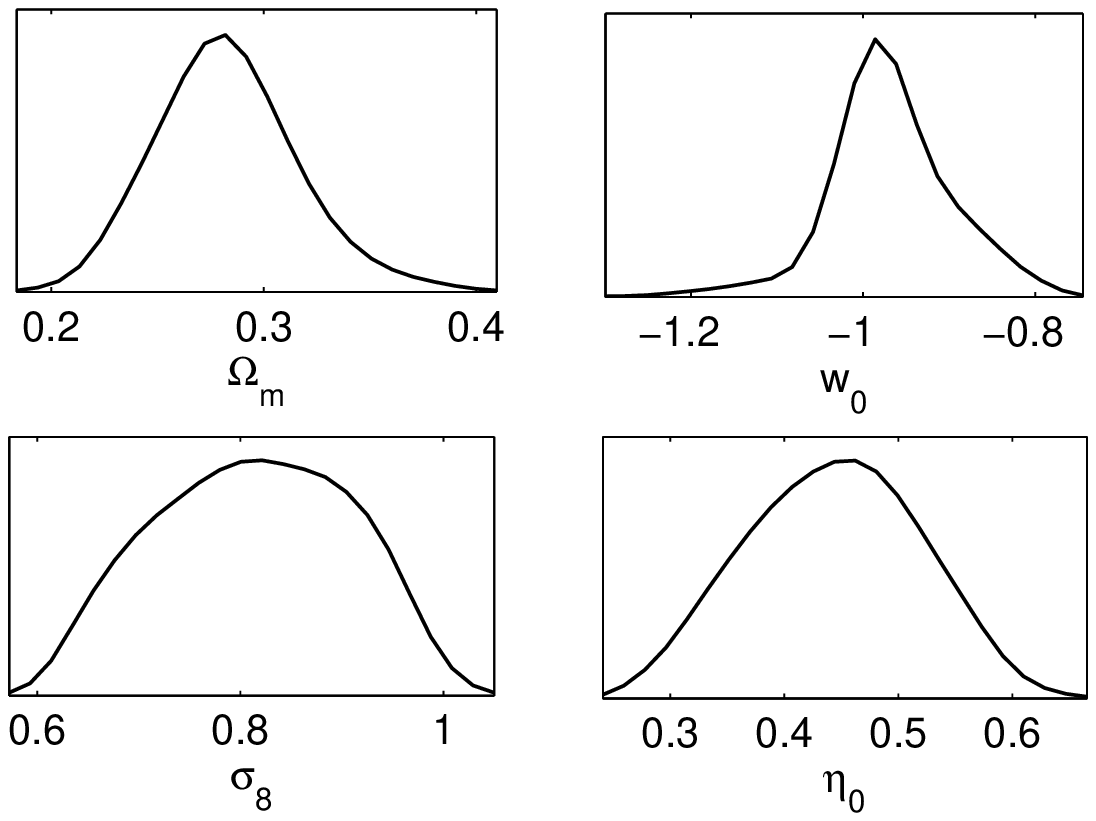}
    }
    \subfigure[]{
     \centering
      \includegraphics[width=7cm]{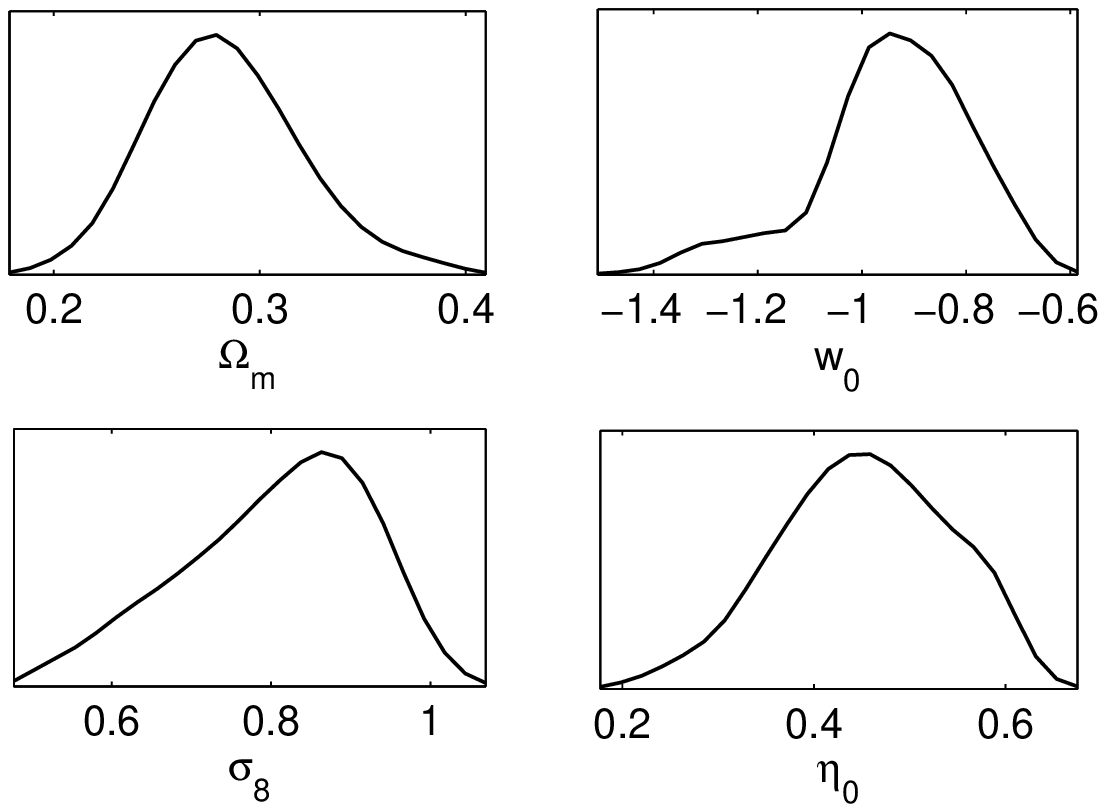}
    }
}
\caption[]{Marginalized pdf's for a few parameters obtained from the analysis of mock cluster data
 of the DES using MCMC methods. {\bf (a)} $w$CDM model with a one parameter scatter; {\bf (b)} $w$CDM 
model with a two parameter scatter}
\label{fig:DES_1D}
\end{figure}

\begin{figure}[ht]
\centering
  \includegraphics[width=15cm]{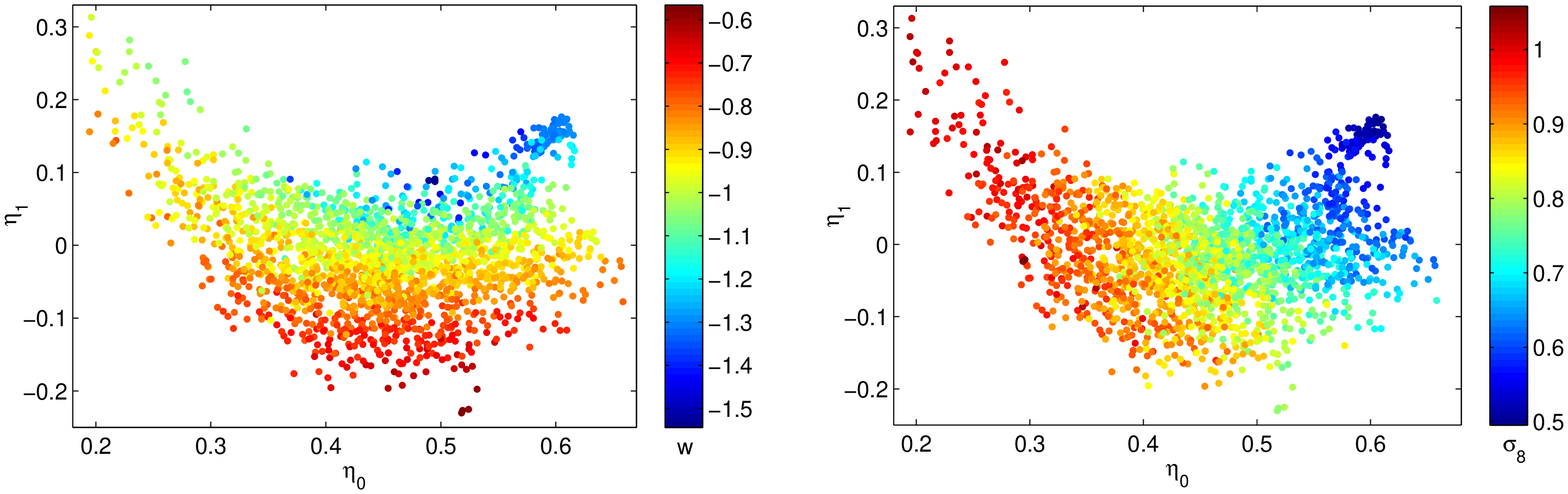}
\caption[]{An analysis of mock data for DES using MCMC methods shows the degeneracy between the scatter parameters $\eta_0$ and $\eta_1$ 
as well as their correlation with a third parameter -- $w_0$ (left) and $\sigma_8$ (right). The dots indicate a few samples 
drawn from the posterior distribution, colored according to the value of the third parameter, $\Omega_m$ or $\gamma$.}
\label{fig:DES_3D}
\end{figure}

\paragraph{DES :}
For a larger optical survey like DES we also examine the effect of introducing $\eta_1$ as the free
 parameter for the redshift dependent scatter. As expected, with an extra parameter the errors on $\Omega_m$ and $\sigma_8$
get larger from 0.025 to 0.030 and 0.070 to 0.113 respectively, for the $\Lambda$CDM model. For the $w$CDM case the
 increase is smaller, 0.033 to 0.037 for $\Omega_m$ and 0.094 to 0.116 for $\sigma_8$. The parameter $w_0$ is
 seen to be degenerate with $\eta_1$, see figure \ref{fig:DES_eta1}; and this causes a significant degradation in the constraints on $w$, from 0.071
 to 0.149, with and without $\eta_1$, respectively. The Fisher results for $w_0$ are always underestimated; 
 for $\sigma_8$ they are either overestimated (up to 50\%) or remain same; while for $\Omega_m$ these are seen to be more
 or less consistent. This may be understood by observing the shape of the peaks in the marginalized single
 parameter pdf's -- for a sharply peaked curve such as in $w_0$, the Fisher results would be underestimated, while 
for a flatly peaked curve, they would be overestimated, see figure \ref{fig:DES_1D}. The Fisher constraints on $\eta_0$ 
are overestimated, but underestimated for $\eta_1$. The marginalized pdf's in this figure show asymmetry as well
 as the presence of extended non-Gaussian tails. The degeneracy of $\sigma_8$ with $\eta_1$ is seen to be much 
weaker than with $\eta_0$, however the degeneracy shape is highly non-Gaussian and skewed, see figure 
  \ref{fig:DES_3D} causing a factor of 1.5--2 discrepancy in $\Delta \eta_1$, in Fisher and MCMC
 forecasts. The above mentioned results are also listed in detail in Table \ref{tab:DES_results}.

\section{Improved constraints from a combination of clusters with CMB+BAO+SNe observations}
We also we consider the effect of adding the results from CMB+BAO+SNe to the $\frac{dN}{dz}$ constraints from cluster surveys.
 We add Gaussian priors of  $\Delta \Omega_m$=0.015, $\Delta w$=0.053 and $\Delta \sigma_8$=0.038 obtained from the WMAP7
 analysis. We find a significant improvement in the cosmological constraints on adding these priors. For the eROSITA and 
WFXT surveys $\Delta \sigma_8$ value can be improved by a factor of 2.5--3 over the WMAP7 constraints. 
The constraints on the dark energy equation of state $w_0$ shrink by half for WFXT and DES surveys, in comparison to 
the results from just CMB+BAO+SNe. These results are also summarized in table \ref{tab:CMB_priors}.


\begin{table}[ht]
\hfil
\begin{tabular}{|c|c|cc|cc|cc|}
\hline survey & CMB &\multicolumn{2}{c|}{+eROSITA} & \multicolumn{2}{c|}{+WFXT} & \multicolumn{2}{c|}{+DES} \\
parameter & +SNe &$\frac{dN}{dz}$+FUP & $\frac{dN}{dz}$+FUP & $\frac{dN}{dz}$+FUP &  $\frac{dN}{dz}$+FUP & $\frac{dN}{dz}$+FUP &  $\frac{dN}{dz}$+FUP\\
          & +BAO &   & +WMAP7 &          & +WMAP7   &         & +WMAP7 \\    
\hline
$\Omega_m$ 		&0.015 & 0.035	& 0.013 & 0.029	& 0.011 & 0.033	& 0.013 \\
$w_0$ 			&0.053 & 0.191	& 0.042	& 0.133	& 0.025 & 0.071	& 0.026 \\
$\sigma_8$ 		&0.038 & 0.034	& 0.011 & 0.031	& 0.013 & 0.094	& 0.035 \\
\hline
\end{tabular}
\hfil
\caption{Improvement in the $1-\sigma$ marginalized parameter constraints 
(from MCMC analysis) from adding clusters to WMAP7
 analysis of CMB+BAO+SNe datasets. All constraints are for the $w$CDM model.}
\label{tab:CMB_priors}
\end{table}

\begin{table}[ht]
\hfil
\begin{tabular}{|c|c|c|r|r|r|r|r|r|}
\hline
parameter & fiducial & mean & \multicolumn{2}{c|}{1-$\sigma$} & \multicolumn{2}{c|}{2-$\sigma$} & \multicolumn{2}{c|}{3-$\sigma$} \\
\cline{4-9}
	  &	     &	    & MCMC & Fisher & MCMC & Fisher & MCMC & Fisher \\
\hline
\hline
	   &         &         & 0.408	& 0.339	& 0.502 & 0.396	& 0.608	& 0.453    \\[-1ex]
\raisebox{1.5ex}{$\Omega_m$} & \raisebox{1.5ex}{0.282}   &  \raisebox{1.5ex}{0.346}  & 0.284  & 0.225 & 0.256 & 0.168 & 0.229 & 0.111    \\
\cline{4-9}
 	   & 	     & 	       & -0.700   & 0.811    & -0.492	   & -0.622   & -0.294   & -0.433  \\[-1ex]
\raisebox{1.5ex}{$w_0$}    & \raisebox{1.5ex}{-1}     & \raisebox{1.5ex}{-0.897} & -1.092   & -1.189    & -1.293	   & -1.378   & -1.557   & -1.567  \\
\cline{4-9}
	   &    &    & 0.802    & 0.846    & 0.823 	   & 0.890     & 0.843    & 0.934     \\[-1ex]
\raisebox{1.5ex}{$\sigma_8$} & \raisebox{1.5ex}{0.802}   &  \raisebox{1.5ex}{0.760}  & 0.718     & 0.758     & 0.667  	   & 0.714      & 0.618     & 0.670     \\
\hline
\end{tabular}
\hfil
\caption{A comparison of 1-$\sigma$, 2-$\sigma$ and 3-$\sigma$ marginalized limits on the cosmological parameters 
 obtained using MCMC and Fisher analysis of the mock cluster data from the Planck SZ survey 
for the $w$CDM model. 
} 
\label{tab:Planck_limits}
\end{table}

\begin{figure}[ht]
\centering
  \includegraphics[width=15cm]{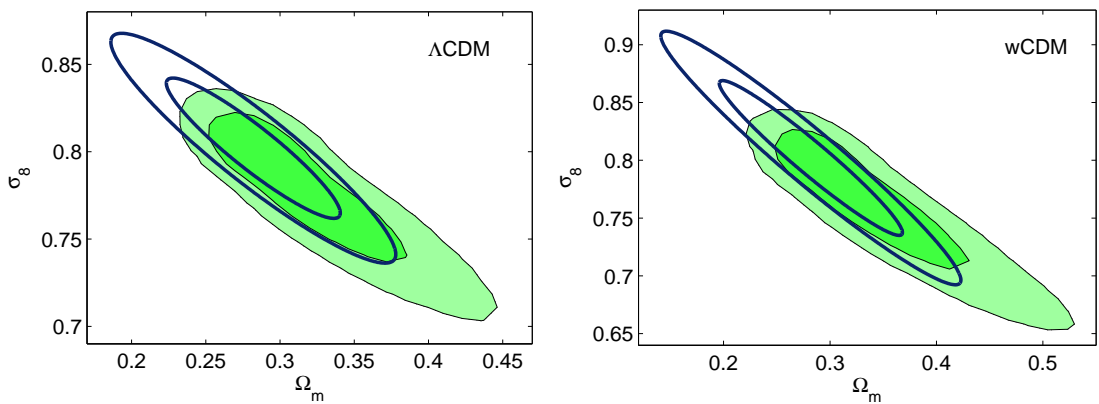}  
\caption[]{The dark (light) green regions indicate the projected 1-$\sigma$ (2-$\sigma$) confidence regions obtained 
from the analysis of mock cluster data for the Planck SZE survey using MCMC methods for the $\Lambda$CDM and $w$CDM cosmological 
models in the $\Omega_m - \sigma_8$ plane. The blue ellipses are the corresponding constraints from a Fisher analysis.}
\label{fig:Planck_om_s8}
\end{figure}

\section{Improved constraints from combining multiple cluster surveys}
\label{subsec:synergy}
In the previous section we showed how having clusters together with CMB, SNe \& BAO leads to percent level constraints on key cosmology parameters. In this section we point out a way to have tight cosmological constraints, than possible in section \ref{sec:results}, but using cluster datasets only.
The surveys that we have discussed till now along with SZE survey (like Planck) will have datasets which will include clusters common in overlapping  regions of the sky. A subset of these clusters can be used innovatively to constrain cosmological constraints from using number counts only from any
particular dataset \cite{Satej2010b}. However, we show below, that due to the different degeneracy directions among cosmological and cluster parameters in each of these surveys, a joint analysis of $\frac{dN}{dz}$ from any two surveys will greatly increase the constraining power of clusters as cosmological probes.
We show this for three sets: (i) optical (DES) $+$ x-ray (WFXT), (ii) optical (DES) $+$ SZE (Planck), and (iii) x-ray (eROSITA) $+$ SZE (Planck). Before proceeding further, we look at the prospects from Planck clusters.

\paragraph{Planck Cluster Cosmology :}
With an estimated yield of $\sim 2000$ clusters,  Planck gives the following constraints -
 $\Delta \Omega_m$ = 0.044 and 0.064, while $\Delta \sigma_8$ = 0.028 and 0.041 for the $\Lambda$CDM and $w$CDM models
 respectively with $\Delta w_0$ = 0.199 in the latter case. We find that fairly
 good agreement exists between the Fisher and MCMC estimates for Planck. However, it is interesting to take note of the 
fact that the 2-dimensional projected likelihoods obtained using Fisher in the $\Omega_m - \sigma_8$ plane show a 
substantial offset as compared to the MCMC likelihoods, see figure \ref{fig:Planck_om_s8}. This is seen 
in both the $\Lambda$CDM as well as $w$CDM models, and arises due to the asymmetric distribution of the 
posterior likelihoods, especially so in the plane of these two parameters. 
Table \ref{tab:Planck_limits}
shows a comparison of 1-$\sigma$, 2-$\sigma$ and 3-$\sigma$ limits on the marginalized parameter 
constraints obtained using MCMC and Fisher analysis Planck SZ survey for the $w$CDM model.

It is worth emphasizing again at this point that the surveys considered in this work give rise to very 
different degeneracy directions in the parameter space. If these degeneracies could be broken through a joint analysis 
of various cluster datasets, we would obtain significantly tighter cosmological constraints {\it using clusters alone}.
 We now give a few examples to show how, and also to indicate the extent to which this could be possible. A detailed
 report of such improvements in the constraints from such a joint analysis will be reported in a future work.

\begin{figure}[ht]
\centering
  \includegraphics[width=15cm]{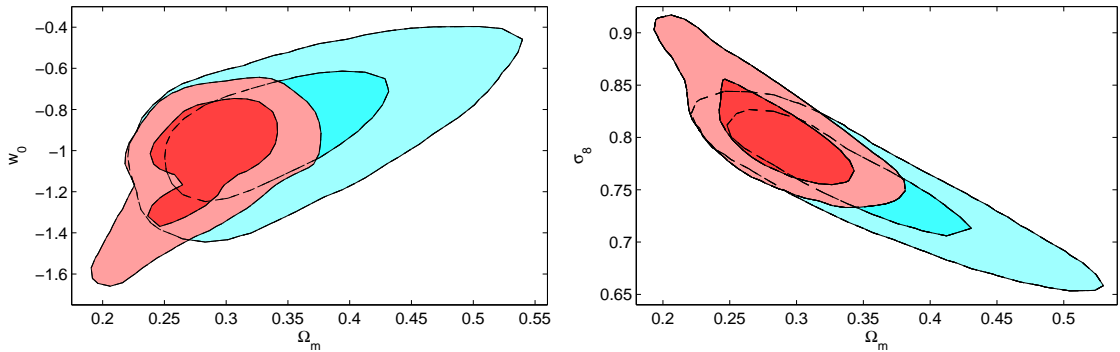}
\caption[]{The dark(light) regions indicate the projected 1-$\sigma$ (2-$\sigma$) confidence regions obtained from the analysis 
of mock cluster data using MCMC methods for a $w_a$CDM model. The cyan (red) regions correspond to constraints from the Planck (eROSITA)
 survey.}
\label{fig:eROSITA_Planck}
\end{figure}

\begin{figure}[ht]
\centering
  \includegraphics[width=15cm]{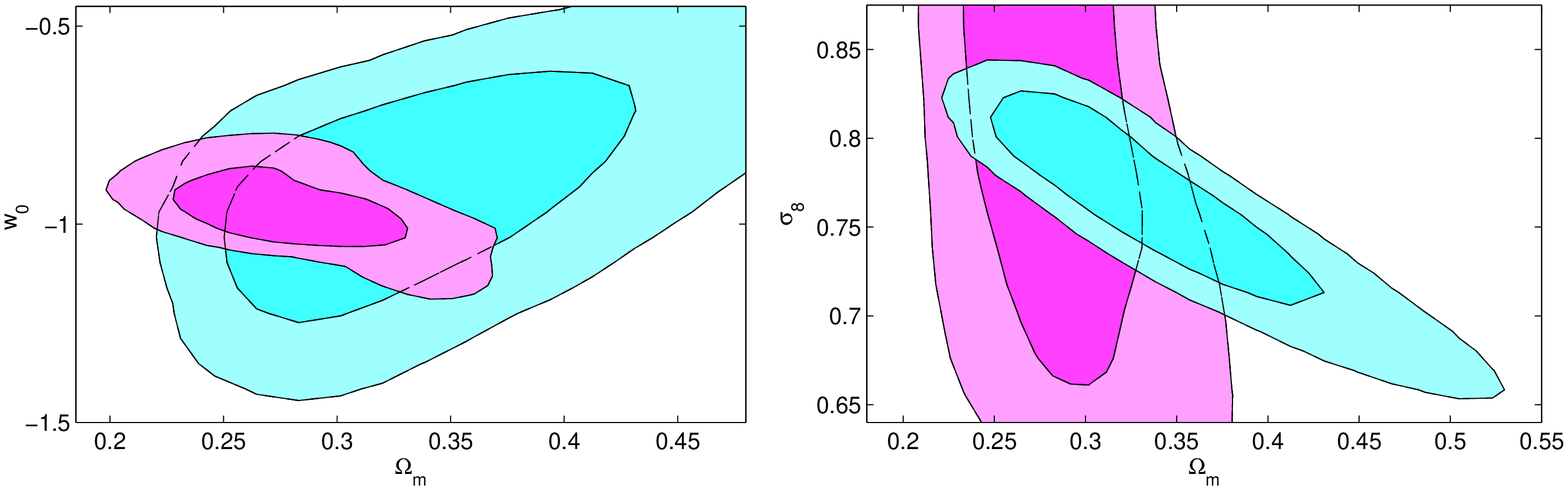}
\caption[]{The dark(light) regions indicate the projected 1-$\sigma$ (2-$\sigma$) confidence regions obtained from the analysis 
of mock cluster data using MCMC methods for a $w_a$CDM model. The cyan (magenta) regions correspond to constraints 
from the Planck (DES) survey.}
\label{fig:Planck_DES}
\end{figure}

\begin{figure}[ht]
\centering
  \includegraphics[width=15cm]{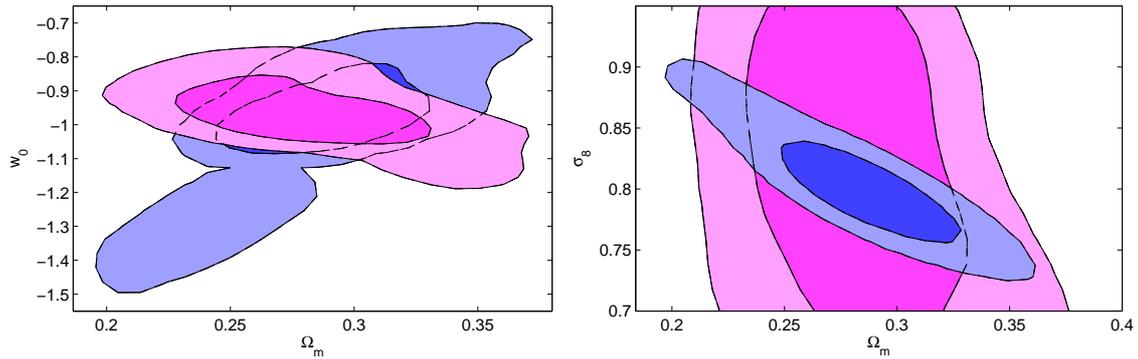}
\caption[]{The dark(light) regions indicate the projected 1-$\sigma$ (2-$\sigma$) confidence regions obtained from the analysis 
of mock cluster data using MCMC methods for a $w_a$CDM model. The blue (magenta) regions correspond to constraints 
from the WFXT (DES) survey.}
\label{fig:WFXT_DES}
\end{figure}
 
The degeneracy directions of the parameters in the $\Omega_m - w$ and $\Omega_m - \sigma_8$ planes for Planck 
and eROSITA surveys are parallel, however the posterior likelihood distributions are seen to be 
asymmetric with the tails being exactly in opposite directions, see figure \ref{fig:eROSITA_Planck}. Each of these surveys individually shows large 
degeneracy, however the overlap region is much smaller.

Next, the optical survey DES and the SZ survey Planck seem to be exactly complimentary. As seen before, the DES is 
expected to give only weak constraints on the parameter $\sigma_8$ due to a large scatter, but at the same time gives 
relatively strong constraints on both $\Omega_m$ as well as the dark energy parameter $w$. It turns out that  
Planck would have an almost orthogonal degeneracy with DES, see figure \ref{fig:Planck_DES}, which could kill the degeneracy 
 to give much better joint constraints.

Finally, we consider the benefits of a joint WFXT and DES analysis. Again, we find that degeneracy directions 
are almost orthogonal between the two surveys, see figure \ref{fig:WFXT_DES}. The WFXT survey gives somewhat 
extended posterior distribution for the dark energy parameter $w$, beyond the 1-$\sigma$ region. Using these 
surveys in combination would produce better 2-$\sigma$ constraints on the parameter $w_0$.

\section{Discussions}
\label{sec:discussions}
In our assumptions made to model future galaxy cluster surveys, we followed a simple
recipe for self-calibration using a mass follow-up. The parameter degeneracies and also the
 constraints certainly depend on the self-calibration methods used to improve upon the
 cosmological constraints. In another work \cite{Satej2010b}, we examine the
 dependence of degeneracies for SZ surveys like ACT/SPT in the $\Omega_m$-$w_0$
 and $w_0$-$w_a$ plane with various choices of calibration (see Fig. 2)
 such as distance measurements from supernovae type Ia, distances from joint X-ray+SZ observations
 of clusters and mass follow-up's with different errors on mass measurements. In \cite{Satej2010a} we
 propose a different method of self-calibration, using a combination of deep and wide SZ surveys to 
obtain significantly tighter cosmological constraints. Here we use the fact that the deep and wide surveys
have slightly different degeneracies, and we show that it is possible to get competitive constraints
 without the need for a costly mass follow-up.

While estimating the parameter constraints $\Delta \theta_i$ from the Fisher technique
 it is important to check for numerical stability. Numerical instability may arise from - 
the numerical derivatives not having converged; or the Fisher matrix ${\mathcal F}$ being 
unstable under inversion. The later occurs especially when the eigenvalues of ${\mathcal F}$
 span a large range of values (This may be avoided to some extent in practice by 
re-parametrizing $\theta_i$ so that $\Delta \theta_i$ for different parameters are not
 too different in magnitude.). For a near-singular ${\mathcal F}$, methods of singular value 
decomposition should be used for the inversion. It is preferable to compute the 
numerical derivative symmetrically about a given value, i.e. for small $h$,
$\frac{\partial f}{\partial \theta} \approxeq \frac{f(\theta + h/2)-f(\theta + h/2)}{h}$. At
 the same time $h$ should not be chosen to be too small, such that the changes in the numerator
 $\Delta f$ become comparable to the accuracy in computing the function $f$. This may give rise to
 artificially inflated values in the parameter constraints.

In our analysis we find that the degeneracies of the parameter $w_a$ with other parameters often shows
 different directions in Fisher and MCMC analysis, for e.g. see Fig. \ref{fig:eROSITA_discrep} and \ref{fig:WFXT_DE_EOS}.
In our view the discrepancy may be explained in the following manner: The direction of the Fisher ellipse is
 related to the ratio of the marginalized errors on the parameters. Since the discrepancy between Fisher and MCMC
 (marginalized) constraints on the parameter $w_a$ are very different, it should not be surprising
 that the degeneracy directions are also different. For example, in the case of eROSITA Fisher, the errors
 from Fisher and MCMC differ by factors of ~2.2 and ~4.2 for the parameters $w_0$ and $w_a$ respectively.
This causes the conflict when comparing with the degeneracy as seen from MCMC.

Our MCMC analysis is based on the data collected from 6-7 independently run chains. We use the
 convergence statistic R-1 = variance(chain mean)/mean(chain variance) \cite{Heavens_stats}, 
computed for the second half of the set of seven independently evolved MCMC chains. For the 
Metropolis sampling we rescale the proposal matrix further in order to get an acceptance rate of
 20-25 \%. For a higher acceptance rate the chain keeps jumping around and is more likely to land
 in regions of low probability, while for a low acceptance rate the chains mix very slowly; in both
 the cases convergence is slow. For good convergence, R-1 $\lesssim$ 0.03. We find that our analysis
 requires between 300,000 - 10,000,000 points sampled during the MCMC exploration of parameter space
 for the chains to be well converged. Sometimes, the convergence of chains can be improved by
 annealing the chains, i.e. instead of sampling from the distribution $\mathcal{L}$, one uses 
${\mathcal{L}}^{1/T}$. Here the parameter $T$ plays the role similar to temperature, in the sense of
 broadening the distribution for $T>1$ and vice-versa. This is especially helpful for exploring the tails of
 distributions, discovering other local minima, and for getting more robust high-confidence error bars.
The parameter likelihoods must be adjusted accordingly (or cooled) during the analysis to get back 
the correct target pdf. We find that a mild heating to  $T= 1.5-2$ can boost the convergence significantly.
 However, heating to higher $T$ may cause the chains to approach the limits set by allowed values of parameters,
 for e.g. $0 < \Omega_m < 1$, which may cause an artificial cut-off in the distribution on cooling back the chains.
 In addition, it is also useful to thin the chains by factors of 10-100 in order to reduce the auto-correlations
 in parameter values of the chains.

Finally, in our MCMC analysis we did not consider the effect of scatter in both $\frac{dN}{dz}$ and the follow-up data,
 as it is not possible to introduce scatter in the the simple Fisher analysis.
 However, real data is always expected to contain some scatter and the best fit parameter
 values are seen to move away from the fiducial values as scatter brings in some extra
 freedom in the fitting of data. 

\section{Conclusion}
\label{sec:conclusion}
We have made a detailed study of the cosmological forecasts for  upcoming cluster surveys showing that  
 these cluster surveys would be able to place strong constraints on cosmology even with 
a simple mass follow-up of about a 100 clusters. In comparing forecasts from Fisher estimates to those from MCMC analysis, we strongly advocate the use of full MCMC likelihood in forecasting parameter constraints, especially on those related to dark energy. We find that the Fisher estimates are reliable only
 for predicting the constraints on minimal cosmological models. In many cases the results from the two methods 
are seen to diverge by factors of 1.5 -- 2. We show that the Fisher estimates can completely fail to correctly forecast the 
constraints as well as the degeneracy directions for cosmological models with a dark energy whose equation of 
state $w(z)$ evolves with redshift.
For some surveys we find an asymmetric posterior distribution for the cosmological parameters. For example, 
the MCMC analysis of the mock cluster data expected from the Planck survey shows a large offset compared 
to the Fisher ellipses. In the case of optical cluster surveys we see that, a large unknown scatter in the
mass-proxy relation degrades the constraints on $\sigma_8$. $\Delta \sigma_8$ further degrades if the 
scatter is also allowed to be redshift dependent. This occurs due to the large degeneracy between 
$\sigma_8$ and scatter. The redshift dependent scatter is seen to weaken the constraints on $w_0$ by
 a factor of 2 or so. The presence of scatter, especially a redshift dependent one, is seen to give 
a skewed distribution for the posterior likelihoods as well as non-Gaussian tails in the marginalized 
single parameter pdf's. Breaking the degeneracies between scatter and $\sigma_8$ would be 
particularly important for future cluster surveys in optical (where the scatter is estimated to be 
large) in order to get competitive constraints with other probes. 
In the end we show through some specific 
examples that even though the parameter constraints obtained from the analysis of some of the 
cluster survey data are relatively weak to begin with, a joint analysis of the datasets from two 
(or more) cluster datasets can be  effectively utilised to kill the near orthogonal parameter degeneracies
 occurring from these surveys to obtain significantly tighter cosmological constraints.

In general we observe that the discrepancy between the two forecasting methods increases with the 
introduction of additional parameters like $w_a$ and $\eta_1$ for the redshift dependence of dark
 energy and scatter respectively. Most of the forecasts on cluster cosmology presented in the literature have used the 
Fisher matrix analysis, mainly due to its simplicity. In our view, such forecasts, even though 
being agreeable with MCMC in many cases need to be accepted with a fair bit of caution. This is 
especially important when additional parameters relevant to cosmology or cluster physics are 
introduced in the analysis of data -- parameters for which we are {\it ab initio} unaware about the Gaussianity
of the likelihood distribution. Moreover, an accurate knowledge of parameter degeneracies, available from MCMC, is very 
 important for joint analysis of datasets from two or more surveys to obtain tighter 
constraints on cosmological and cluster physics. 

\section{Acknowledgements}
The authors would like to thank the department of Theoretical Physics at TIFR for the use of the
 computing cluster {\it Brood} on which most of this work was carried out. We also acknowledge
 the use of {\it GetDist} program from the publicly available package {\it COSMOMC} to analyze our MCMC chains.
S.K. would like to thank Joe Mohr, Thomas Reiprich, Matthias Bartelmann, Steen Hansen and Stefano Andreon
 for many discussions. The anonymous referee is thanked for providing several useful comments and suggestions
 that have helped in improving the presentation of this work.

\providecommand{\href}[2]{#2}\begingroup\raggedright\endgroup

\end{document}